\documentclass[pra,a4paper,twocolumn,superscriptaddress,longbibliography,nofootinbib]{revtex4-2}
\usepackage[utf8]{inputenc}
\usepackage{amssymb}
\usepackage{amsmath}
\usepackage{amsfonts}
\usepackage{amsthm}
\usepackage{amsbsy}
\usepackage{bm}
\usepackage{bbold}

\usepackage{graphicx}
\usepackage[dvipsnames]{xcolor}
\usepackage{multirow}
\usepackage{float}
\usepackage{comment}
\usepackage{ulem}
\usepackage{relsize}
\usepackage{cmap}
\usepackage{physics}
\usepackage[colorlinks]{hyperref}

\DeclareMathOperator{\sgn}{sgn}

\DeclareMathOperator{\arcsinh}{arcsinh}

\DeclareMathOperator{\so}{\textsf{s}_\textsf{0}}
\DeclareMathOperator{\sx}{\textsf{s}_\textsf{1}}
\DeclareMathOperator{\sy}{\textsf{s}_\textsf{2}}
\DeclareMathOperator{\sz}{\textsf{s}_\textsf{3}}
\DeclareMathOperator{\sj}{\textsf{s}_\textsf{j}}

\DeclareMathOperator{\li2}{li_2}
\DeclareMathOperator{\llangle}{\langle\!\langle}
\DeclareMathOperator{\rrangle}{\rangle\!\rangle}
\DeclareMathOperator{\Llangle}{\Bigl \langle\!\!\Bigl \langle}
\DeclareMathOperator{\Rrangle}{\Bigr \rangle\!\!\Bigr\rangle}
\DeclareMathOperator{\DD}{\mathcal{D}\!\!\mathcal{D}}

\newcommand{\DOS}{{LDoS}}
\newcommand{\NLSM}{{NL$\sigma$M}}
\begin{document}
	
	\title{Boundary multifractality in the spin quantum Hall symmetry class with interaction}

        \author{S. S. Babkin}
        
        \affiliation{Institute of Science and Technology, Am Campus 1 3400 Klosterneuburg, Austria} 
        
\author{I. S. Burmistrov}
	
	\affiliation{\hbox{L.~D.~Landau Institute for Theoretical Physics, acad. Semenova av. 1-a, 142432 Chernogolovka, Russia}}
	
        \affiliation{Laboratory for Condensed Matter Physics, HSE University, 101000 Moscow, Russia}

	\date{\today} 
	
\begin{abstract}
Generalized multifractality characterizes system size dependence of pure scaling local observables at Anderson transitions in all ten symmetry classes of disordered systems. Recently, the concept of generalized multifractality has been extended to boundaries of critical disordered noninteracting systems. Here we study the generalized boundary multifractality in the presence of electron-electron interaction, focusing on the spin quantum Hall symmetry class (class C). Employing the two-loop renormalization group analysis within Finkel'stein nonlinear sigma model we compute the anomalous dimensions of the pure scaling operators located at the boundary of the system. We find that generalized boundary multifractal exponents are twice larger than their bulk counterparts. Exact symmetry relations between generalized boundary multifractal exponents in the case of noninteracting systems are explicitly broken by the interaction.
\end{abstract}

	\maketitle
	%%%%%%%%%%%%%%%%%%%%%%%%%%%%%%%%%%%%

\section{Introduction}

A fascinating example of a quantum phase transition in a free fermion system is 
Anderson transition \cite{Anderson58}. This transition is controlled by disorder and separates metallic and insulating phases. Additional boost to studies of Anderson transition is provided by the fact that some Anderson transitions occur between distinct topological (insulating) phases, e.g. integer quantum Hall plateau-plateau transitions. An intriguing feature of Anderson transition is strong mesoscopic fluctuations of electron wave functions at criticality \cite{Wegner1980,Castellani1986}. Consequently, the disorder averaged moments of the local density of states ({\DOS}) demonstrate {\it pure} power-law scaling with the system size, $\langle \rho^q\rangle {\sim} L^{{-}x_{\textsf{(q)}}}$. Here values
of
the {\it multifractal} exponents $x_{\textsf{(q)}}$ depend on a symmetry class of considered random Hamiltonian (see Refs. \cite{Mirlin2000,EversMirlin} for a review). 

There are much more pure scaling observables in addition to the moments of {\DOS}~\cite{Wegner1986}. They can be expressed in terms of disorder averages of specific combinations of wave functions \cite{Gruzberg2013,Karcher2022,Karcher2022b,Karcher2023}. The corresponding set of multifractal exponents $x_\lambda$, termed as {\it generalized multifractality}, is unique characteristic of Anderson transition in each symmetry class. Exponents $x_\lambda$ are related by symmetry relations specific for each symmetry class ~\cite{Mirlin2006,Gruzberg2011,Gruzberg2013}.

Recently, it has been established that statistics of wave functions at surface ($s$) of a system undergoing bulk Anderson transition is different from the statistics in the bulk \cite{Subramaniam2006, mildenberger2007, Subramaniam2008, Evers2008,Obuse2008}. In particular, the scaling of the {\DOS} moments at the boundary is given as $\langle\rho^{q}(\bm{r}{\in} s)\rangle{\sim} L^{{-}x^{(s)}_{\textsf{(q)}}}$, with $x^{(s)}_{\textsf{(q)}}{\neq}x_{\textsf{(q)}}$. In Ref. \cite{Babkin2023} the theory of generalized multifractality  has been extended to boundaries of critical systems. 

%For example, the surface multifractality is expected to be of relevance for experiments on scanning tunneling microscopy of the surface of a magnetic semiconductor Ga$_{1-x}$Mn$_x$As near the three-dimensional metal-insulator transition \cite{richardella2010}. 
%Furthermore, it was shown that, for a certain range of $q$,  surface effects have a dominant contribution to the multifractality of the entire system (including bulk and surface) \cite{Subramaniam2006}. 

The picture of generalized multifractality at Anderson transitions has recently been fully supported by numerics in symmetry classes A, C, AIII, AII, D, and DIII \cite{Karcher2021,Karcher2022,Karcher2022b,Karcher2023a,Karcher2023}. However, multifractality is of relevance for experiments as well. Light waves spreading in an array of dielectric nanoneedles demonstrated multifractal behavior in experiments reported in Ref.  \cite{Mascheck2012}. Multifractal behavior of ultrasound waves was observed while they propagated through a system of randomly packed Al beads \cite{Faez2009}. In the experiment  \cite{Richardella} the electron {\DOS} was measured by scanning tunneling microscopy on a surface of diluted magnetic semiconductor Ga$_{1{-}x}$Mn$_x$As. While it was tuned through bulk Anderson transition multifractal signatures in {\DOS} has been measured. In the experiment on Ga$_{1{-}x}$Mn$_x$As, the surface multifractality was presumably observed. Multifractal behavior of {\DOS} amplitude has been recently measured in weakly disordered superconducting state in the stripped incommensurate phase of monolayer Pb/Si(111) \cite{Lizee2023}.

Multifractality is responsible for many nontrivial physical effects. It was shown \cite{FeigelmanYuzbashyan2007,FeigelmanCuevas2010,Burmistrov2012,Burmistrov2015b,DellAnna,Burmistrov2021,Andriyakhina2022,Nosov2023} that multifractal correlations effectively increase electron-electron attraction and, thus, lead to strong enhancement of superconducting transition temperature and the superconducting gap at zero temperature. Moreover, it was found that multifractality is responsible for log-normal distribution of the superconducting order parameter \cite{FeigelmanCuevas2010,Fan2020,Stosiek2020} and {\DOS} \cite{Burmistrov2021,Stosiek2021,Andriyakhina2022} in dirty superconducting films. Multifractal correlations result in instabilities of surface states in topological superconductors \cite{Foster2012,Foster2014}. Multifractal behavior of {\DOS} causes strong mesoscopic fluctuations of the Kondo temperature \cite{Kettemann2006,Micklitz2006,Kettemann2007}.  Multifractality affects electron-phonon coupling making cooling of electrons more efficient \cite{Feigelman2019}.  Anderson orthogonality catastrophe is also affected by multifractal properties of wave functions \cite{Kettemann2016}. Multifractality in {\DOS} enhances depairing effect of magnetic impurities on superconducting state in dirty films \cite{Burmistrov2018} and the superconducting {\DOS} around Yu-Shiba-Rusinov states \cite{Babkin2022Y}. 

Recently, it has been suggested that multifractality can serve as a sensitive instrument 
to test critical theories proposed to describe Anderson transitions. Although for each of ten Altland--Zirnbauer symmetry classes effective, long-wave description in terms of a nonlinear sigma model ({\NLSM}) is known (see Ref. \cite{EversMirlin} for a review), Anderson transition occurs typically in strong coupling. Thus Anderson transition criticality is beyond standard treatment of {\NLSM}. A prime example of 
such a situation is the integer quantum Hall plateau transition for which the Wess--Zumino--Novikov--Witten models were conjectured to be an ultimate conformal critical theory \cite{Zirnbauer1999,Kettemann1999,Bhaseen2000,Tsvelik2001,Tsvelik2007,Zirnbauer2019}. It turns out that assumptions of the local conformal invariance and Abelian fusion rules result in parabolic form~\footnote{We note that parabolicity of $x_\lambda$ arises in any dimensionality $d{\geqslant} 2$ in the case of conformal invariance \cite{Padayasi2023}.} of the generalized multifractal exponents $x_{\lambda}$ with a single free parameter only \cite{Bondesan2017,Karcher2021}. However, available numerical computations of multifractal spectrum for the integer quantum Hall plateau transitions demonstrate significant deviations from the exact parabolicity \cite{Obuse2008,Evers2008,Karcher2021}. This makes the theoretical suggestions 
of the Wess--Zumino--Novikov--Witten models  as critical theories for the integer quantum Hall transition to be highly questionable. 

Even more dramatic situation is in superconducting cousin of the integer quantum Hall effect --- the spin quantum Hall effect (class C) \cite{Kagolovsky1999,Senthil1999}. An advantage of the spin quantum Hall transition in $d{=}2$, is that an infinite subset of generalized multifractal exponents is known analitically from exact mapping to the percolation problem \cite{Gruzberg1999,Beamond2002,Mirlin2003,Evers2003,Subramaniam2008,Karcher2022}. The rigorous analytical results serve as a benchmark against numerical computations. Although numerical data for the generalized multifractal spectrum reproduce exact analytical results, they 
demonstrate clear evidence for a violation of parabolicity \cite{Mirlin2003,Puschmann2021,Karcher2021,Karcher2022,Karcher2022b,Karcher2023a,Karcher2023}. Similarly, parabolicity is expected to hold for the surface generalized multifractal exponents in the presence of the local conformal invariance and the abelian fusion. Again, for the class C numerics does not support parabolicity of boundary multifractal exponents but coincides, simultaneously, with exact analytical values of exponents \cite{Babkin2023}. 
These results prove a lack of the local conformal invariance at the spin quantum Hall transition in $d{=}2$.

Electron-electron interaction, being typically, a relevant perturbation modifies the scaling properties of observables at Anderson (or in that case the so-called Mott--Anderson) transitions (see Refs. \cite{Fin,Belitz1994} for a review). Surprisingly, the generalized multifractality exists even in the presence of interaction, i.e. at Mott-Anderson criticality \cite{Burmistrov2013,Burmistrov2015m,Repin2016,Babkin2022}.
In this case the pure scaling operators can be formulated as proper correlations of single particle Green's functions. In particular, the moments of {\DOS} remain pure scaling operators. Although interaction does not change the form of the pure scaling operators (except straightforward  generalization to incorporate a set of Matsubara frequencies) 
it affects the generalized multifractal exponents. In particular, it breaks the symmetry relations between different multrifractal exponents. 

In this paper we develop the theory of the generalized boundary multifractality for the spin quantum Hall symmetry class in the presence of electron-electron interaction. Using Finkel'stein {\NLSM} for class C we compute the anomalous dimensions of 
the pure scaling derivativeless local operators situated near the boundary in the two-loop renormalization group approximation. Surprisingly, within two-loop approximation we find that the anomalous dimensions of pure scaling operators at the boundary and in the bulk differs by a factor $2$. Also interaction breaks the symmetry relations between the generalized surface multifractal exponents in the same way as for the bulk ones. 

Throughout the paper we use terms `surface' and `boundary' interchangeably, as they both  have been used in the previous literature on multifractality. Also we note that in $d$-dimensions, the surface is understood as a $(d{-}1)$-dimensional boundary.	
	
The outline of the paper is as follows. In Sec. \ref{Sec:Formalism} we remind  formalism of Finkel'stein {\NLSM} for class C.  We remind the results for generalized bulk multifractality in the presence of interaction (Sec. \ref{Sec:GSM:B}). The original results for generalized surface multifractality in the presence of interaction are presented in Sec. \ref{Sec:GSM:S}. We end the paper with discussions and conclusions in Sec. \ref{Sec:Final}. The details of computations are given in Appendix.

\section{Finkel'stein {\NLSM} for class C \label{Sec:Formalism}}
	
\subsection{{\NLSM} action}	
	
We start from a brief reminder of the Finkel'stein {\NLSM} for the class C (for more details see Ref. \cite{Babkin2022}). The grand canonical partition function is given as
\begin{equation}
Z=\int D[Q] \exp S, \qquad S=S_{\rm 0} + S_{\rm int},
\label{eq:NLSM}
\end{equation}
where $S_{\rm 0}$ and $S_{\rm int}$ are free and interacting parts of the action.
We note that the action involves also the topological term similar to the class A. However, we omit the topological term in this paper since we focus on the perturbative treatment of the model.
$S_{\rm 0}$ and $S_{\rm int}$ are as follows
\begin{subequations}
\begin{align}
S_{\rm 0} & = -\frac{g}{16} \int_{\bm{r}} \Tr (\nabla Q)^2 + Z_\omega \int_{\bm{r}} \Tr \hat\varepsilon  Q ,
\label{eq:S:0-0}
 \\
S_{\rm int}& =-\frac{\pi T \Gamma_t}{4} \sum_{\alpha,n} 
\int_{\bm{r}} \Tr (I_n^\alpha \bm{\textsf{s}} Q)
 \Tr (I_{-n}^\alpha \bm{\textsf{s}} Q) ,
 \label{eq:S:int}
\end{align}
\end{subequations}
where $\int_{\bm{r}}\equiv \int d^d\bm{r}$ and $T$ stands for temperature. The field variable $Q$ is a Hermitian matrix, $Q^\dag{=}Q$, acting in the $2{\times} 2$ spin space, in the $N_r{\times} N_r$ replica space,  and  in the $2N_m{\times} 2N_m$ space of the Matsubara fermionic energies, $\varepsilon_n{=}\pi T(2n{+}1)$. The matrix $Q$ satisfies the nonlinear local constraint
\begin{equation}
Q^2(\bm{r})=1 
\label{eq:nonlin-const}
\end{equation}  
and obeys the Bogolubov {--} de Gennes symmetry relation
\begin{equation}
Q=  - \bar{Q}, \qquad \bar{Q}= \sy L_0 Q^\textsf{T} L_0 \sy .
\label{eq:symm:C}
\end{equation}
Here superscript $^\textsf{T}$ denotes the matrix transposition operation.
Several matrices introduced above are given as follows
\begin{equation}
\begin{split}
(I_k^\gamma)_{nm}^{\alpha\beta}= &\,  \delta_{n-m,k}\delta^{\alpha\beta}\delta^{\alpha\gamma} \so
,\quad 
\hat \varepsilon_{nm}^{\alpha\beta}=\varepsilon_n \, \delta_{nm}\delta^{\alpha\beta} \so , \\
& (L_0)_{nm}^{\alpha\beta}=\delta_{\varepsilon_n,-\varepsilon_m}\delta^{\alpha\beta} \so ,
\end{split}
\end{equation}
where $\so$ is the $2{\times} 2$ identity matrix in the spin space. The Latin indices represent  Matsubara energies whereas the Greek indices correspond to replica space. The vector $\bm{\textsf{s}}{=}\{\sx,\sy,\sz\}$ is the vector of three nontrivial Pauli matrices
\begin{equation}
\sx = \begin{pmatrix}
0 & 1\\
1 & 0
\end{pmatrix}, \quad
 \sy= \begin{pmatrix}
0 & -i\\
i & 0
\end{pmatrix},  \quad 
\sz = \begin{pmatrix}
1 & 0\\
0 & -1
\end{pmatrix} .
\end{equation}
Nonlinear constraint \eqref{eq:nonlin-const} can be resolved by 
\begin{equation}
Q = \texttt{T}^{-1}\Lambda \texttt{T}, \qquad  \Lambda_{nm}^{\alpha\beta} = \sgn \varepsilon_n \, \delta_{nm} \delta^{\alpha\beta}\so
\label{eq:T-rep}
\end{equation}
Here the rotation $\texttt{T}$ is a unitary matrix satisfying 
\begin{equation}
\texttt{T}^{-1}=\texttt{T}^\dag, \quad  (\texttt{T}^{-1})^\textsf{T} L_0 \sy= \sy L_0 \texttt{T} .
\label{eq:relations:T}
\end{equation}
Parametrization \eqref{eq:T-rep} and condition \eqref{eq:relations:T} fix the target space of the {\NLSM} as $Q{\in} \textrm{Sp}(2N)/\textrm{U}(N)$ where $N{=}2N_rN_m$. We note that one needs to take the limits $N_m{\to} \infty$ and $N_r{\to} 0$.

The {\NLSM} action \eqref{eq:NLSM} involves a bare dimensionless spin conductance $g$, a bare exchange interaction $\Gamma_t$, and a parameter $Z_\omega$, which is responsible for frequency renormalization. 

As we shall see below, in order to extract singular infrared behavior within
the {\NLSM} action it is convenient to add the following regulator into the action \eqref{eq:NLSM}:
 \begin{equation}
S_{\rm h} = \frac{g h^2}{8} \int_{\bm{r}} \Tr \Lambda Q .
\label{eq:S:reg:h}
\end{equation}

We note that the {\NLSM} action \eqref{eq:S:0-0}, \eqref{eq:S:int}, and \eqref{eq:S:reg:h} can be reduced to the {\NLSM} for the class A by breaking spin rotation symmetry from SU(2) down to U(1) such that the $Q$ matrix in the spin space acquires the diagonal form 
\begin{equation}
Q=\begin{pmatrix}
Q_\uparrow & 0 \\
0 & Q_\downarrow 
\end{pmatrix}, 
\qquad  Q_\downarrow = - L_0 Q_\uparrow^\texttt{T} L_0 .  
\end{equation}

\begin{figure}[t]
\centerline{\includegraphics[width=0.38\textwidth]{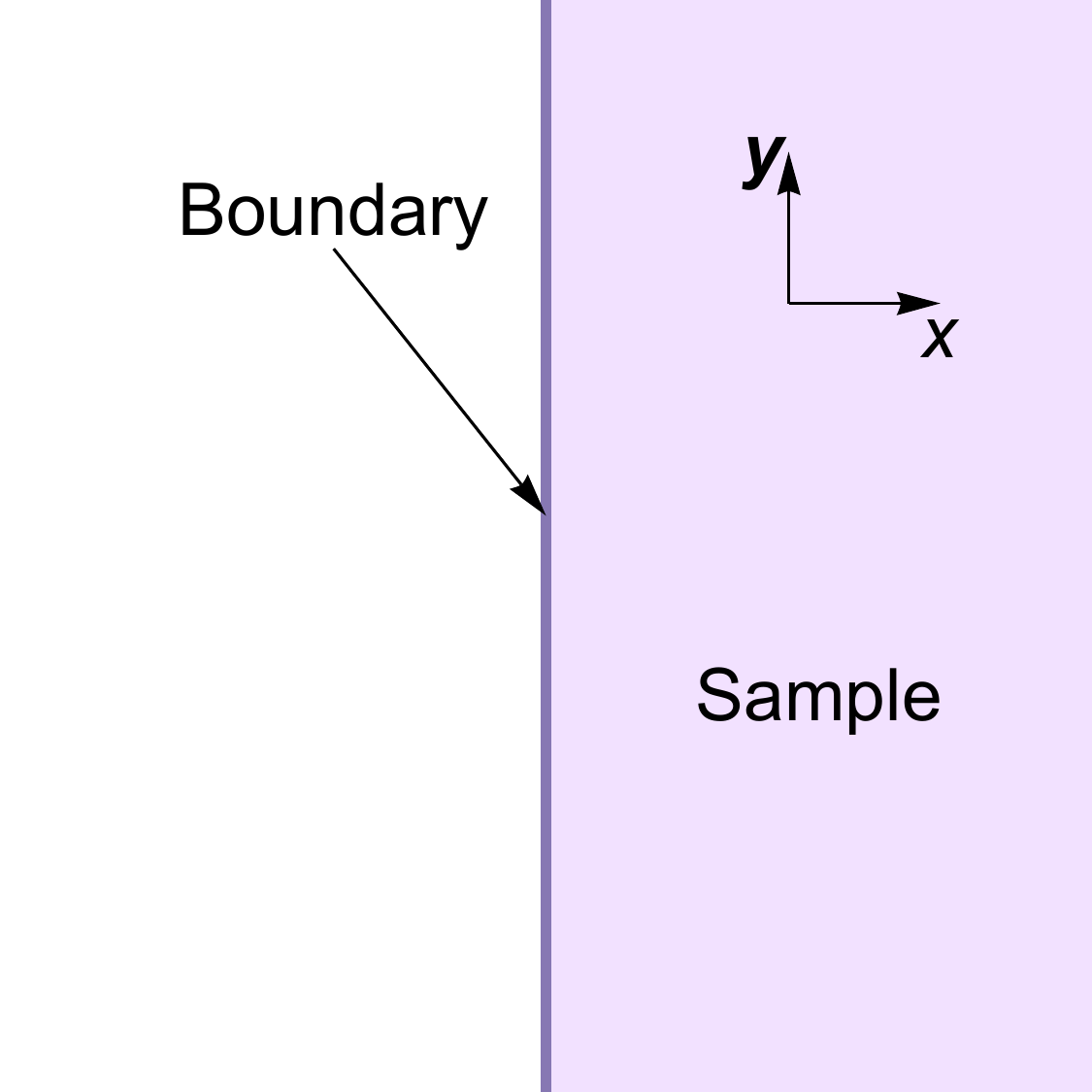}}
\caption{Sketch of the system with a boundary perpendicular to the $x$ axis and situated at at $x{=}0$.}
\label{figure}
\end{figure}

\subsection{Perturbation theory}

In order to proceed we need to develop perturbation theory in $1/g{\ll}1$. Since in this work we are interested in boundary multifractality, we consider a 2D sample with the boundary at $x=0$ (see Fig. \ref{figure}). In what follows, we will employ dimensional regularization method such that we will work in $d{=}2{+}\epsilon$ dimension. We parametrize a $d$-dimensional coordinate vector as $\bm{r}{=}\{x,y_1,{\dots},y_{d-1}\}$.

Also we will use the square-root parametrization of the $Q$ matrix:
\begin{equation}
Q = W+\Lambda \sqrt{1-W^2}, \qquad W = \begin{pmatrix} 
0 & w \\
w^\dag & 0
\end{pmatrix} ,
\end{equation}
where we adopt the following notations: $W_{n_1n_2} {=} w_{n_1n_2}$ and $W_{n_2n_1} {=} w^\dag_{n_2n_1}$ with $\varepsilon_{n_1}{>} 0$ and $\varepsilon_{n_2}{<}0$. Making expansion $w{=}\sum_{\textsf{j}=0}^3 w_\textsf{j} \sj$,
we find that the elements of $w_\textsf{j}$ satisfy the symmetry relations
(cf. Eq. \eqref{eq:symm:C})
\begin{equation}
 (w_\textsf{j})^{\alpha\beta}_{n_1n_2}  = \textsf{v}_\textsf{j}  (w_\textsf{j})^{\beta\alpha}_{-n_2,-n_1}  .
\label{eq:const:C}
\end{equation}
where $ \textsf{v}_\textsf{j} {=} {-}\tr (\sj \sy \textsf{s}_\textsf{j}^\textsf{T} \sy)/2 {=} \{-1,1,1,1\}$. 
	
From the second order expansion of Eq. \eqref{eq:NLSM} in $W$, we find the propagators of Gaussian theory:	
\begin{gather}
{}\hspace{-0.5cm}\Bigl  \langle (w_\textsf{j})^{\alpha\beta}_{n_1n_2}(\bm{r})  (w_\textsf{j}^\dag)^{\mu\nu}_{n_4n_3}(\bm{r^\prime}) \Bigr \rangle 
 {=} \frac{2}{g} 
 \Bigl [\Bigl (\delta^{\alpha\nu}\delta^{\beta\mu}\delta_{n_1n_3}\delta_{n_2n_4}
\notag \\
{+} \textsf{v}_\textsf{j}\delta^{\alpha\mu}\delta^{\beta\nu}\delta_{n_1,{-}n_4}\delta_{n_2,{-}n_3}  \Bigr ) \hat{\mathcal{D}}(i\omega_{n_{12}};\bm{r},\bm{r^\prime})
\notag \\
{-} \frac{4\pi T\gamma}{D}(1{-}\delta_{\textsf{j}0})\delta^{\alpha\nu}\delta^{\beta\mu}\delta^{\alpha\beta}
\delta_{n_{12},n_{34}}\widehat{\DD^t}(i\omega_{n_{12}};\bm{r},\bm{r^\prime}) \Bigr ]  ,
\label{eq:prop:full}
\end{gather}
where we denote $\omega_{n_{12}}{=}\varepsilon_{n_1}{-}\varepsilon_{n_2}$ and $n_{12}{=}n_1{-}n_2$. Next,  $D{=}g/(4Z_\omega)$ and  $\gamma{=}\Gamma_t/Z_\omega$ are a bare diffusion coefficient and a dimensionless interaction strength, respectively.
Diffuson and diffuson dressed by interaction via ladder resummation are given as
\begin{subequations}
\begin{align}
\hat{\mathcal{D}}(i\omega_{n_{12}};\bm{r},\bm{r^\prime}) = \sum_{s=\pm} 
\mathcal{D}(i\omega_{n_{12}};x-s x^\prime,\bm{y}-\bm{y^\prime}) ,\label{eq:diffusons:bound} \\
\hat{\mathcal{D}}^t(i\omega_{n_{12}};\bm{r},\bm{r^\prime}) = \sum_{s=\pm} 
\mathcal{D}^t(i\omega_{n_{12}};x-s x^\prime,\bm{y}-\bm{y^\prime}) ,
\label{eq:diffusons:bound:t}
\end{align}
\end{subequations}
where $\bm{y}=\{y_1,\dots,y_{d-1}\}$. Here $\mathcal{D}(i\omega_{n};x,\bm{y})$ and $\mathcal{D}^t(i\omega_{n};x,\bm{y})$ correspond to the diffusons in an infinite sample ($\int_{\bm q}\equiv \int d^d\bm{q}/(2\pi)^d$),
\begin{gather}
\mathcal{D}^{/t}(i\omega_{n};x,\bm{y})=\int_{\bm q} \mathcal{D}^{/t}_q(i\omega_n)
e^{i q_x x + i \bm{q}_{\|} \bm{y}}
\end{gather}
 with the standard momentum representation
\begin{subequations}
\begin{align}
\mathcal{D}_q(i\omega_n) & =\Bigl [ q^2+h^2+\omega_n/D\Bigr ]^{-1}, \label{eq:prop:def:a}\\
\mathcal{D}^t_q(i\omega_n) & =\Bigl [ q^2+h^2+(1+\gamma)\omega_n/D\Bigr ]^{-1} .
\label{eq:prop:def:b}
\end{align}
\end{subequations}
Also we introduced the following notation
\begin{align}
\widehat{\DD^t}(i\omega;\bm{r},\bm{r_1}) & {=}\!\!
\int^\prime \!\!dx_{2}\! \int \!d^{d{-}1}\bm{y_2}  \hat{\mathcal{D}}(i\omega;\bm{r},\bm{r_2}) %\notag \\\times
\hat{\mathcal{D}}^t(i\omega;\bm{r_2},\bm{r_1}) 
\notag \\
 {=} 
 & \!\int_q \DD^t_q(i\omega) \sum_{s{=}\pm} e^{iq_x(x{-}s x_1){+}i\bm{q_{\|}}(\bm{y}{-}\bm{y_1})}. 
\end{align}
Here the `prime' sign on the integral indicates that we integrate over $x_2{>}0$. Also we introduced the short-hand notation: $\DD^t_q(i\omega_n)\equiv \mathcal{D}_q(i\omega_n)\mathcal{D}^t_q(i\omega_n)$.

{\NLSM} action, see Eqs. \eqref{eq:S:0-0}, \eqref{eq:S:int}, and \eqref{eq:S:reg:h}, is subjected to renormalization. Within one-loop order (the lowest order in $t=1/(\pi g)$), the renormalized parameters (denotes by `prime' signs) are given as (for system without the boundary)
\begin{gather}
h^{\prime 2} = \frac{g h^2 Z}{g^\prime} =
h^2 \Bigl [1 -\frac{b t h^{\epsilon}}{\epsilon}\Bigr ] , \qquad 
g^\prime = g \Bigl [1 + \frac{a_1 t h^{\epsilon}}{\epsilon} \Bigr ] ,\notag \\
\frac{Z_\omega^\prime}{Z_\omega}=\frac{\Gamma_t^\prime}{\Gamma_t}=  1 + (1-3\gamma)\frac{t h^{\epsilon}}{\epsilon} , \notag \\
a_1 = \textsf{v}/2 +6 f(\gamma), \qquad b = 3 \ln(1+\gamma)+6f(\gamma) .
\label{eq:hprime:ren}
\end{gather}
Here we introduced $\textsf{v}{=}\sum_{j=0}^3 \textsf{v}_\textsf{j}{\equiv} 2$. The above results can be rewritten in the form of the one-loop renormalization group equations (with usage of the minimal subtraction scheme \cite{Amit-book}), \begin{subequations}
\begin{align}
\frac{d t}{d\ell} & = -\epsilon t + \bigl [{\textsf{v}}/{2}+6f(\gamma)\bigr ] t^2 + O(t^3),\\
\frac{d \gamma}{d\ell } &  =  0+ O(t^2) , \\
\frac{d \ln Z_\omega}{d\ell } & = - ({\textsf{v}}/{2}-3\gamma)t  + O(t^2) , \\
\frac{d\ln Z}{d\ell} &  = -\bigl [{\textsf{v}}/{2}- 3 \ln(1+\gamma)\bigr ]t+ O(t^2) .
\end{align}
\label{eq:RG:one-loop}
\end{subequations}
Here $\ell{=} \ln 1/h^\prime$ stands for the logarithm of the infrared length scale which is just a system size at $T{=}0$. At finite temperature the infrared scale is set by the temperature length ${\sim}\sqrt{D/T}$. We note that Eqs. \eqref{eq:RG:one-loop} have been derived in Refs.\cite{Jeng2001a,Jeng2001,DellAnna2006,Liao2017,Babkin2022} by various techniques.

\section{Generalized bulk multifractality\label{Sec:GSM:B}}

We start from reminder of generalized multifractality in the bulk for class C reported in Ref. \cite{Babkin2022}. 
An operator without derivatives which involves the number $q$ of  $Q$ fields can be constructed as follows \cite{Repin2016, Babkin2022}.  We introduce the quantity
\begin{gather}
\mathcal{K}_q(E_1,\dots,E_q) = \frac{1}{4^{q}} \sum_{p_1,\dots, p_q =\pm} \left ( \prod\limits _{j=1}^q p_j\right )
\notag \\
\times
\mathcal{P}_q^{\alpha_1,\dots,\alpha_q;p_1,\dots,p_q}(E_1,\dots,E_q)  ,
 \label{eq:Kq}
 \end{gather}
depending on the set $\{E_1,\dots,E_q\}$ of real energies.
The correlation function $\mathcal{P}_q^{\alpha_1,\dots,\alpha_q;p_1,\dots,p_q}(E_1,\dots,E_q)$ can be obtained from its Matsubara counterpart $P_q^{\alpha_1,\dots,\alpha_q}(i\varepsilon_{n_{1}},\dots,i\varepsilon_{n_{q}})$ by the analytic continuation to the real frequencies:  $\varepsilon_{n_j} {\to} E_j{+}i p_j 0^+$. The corresponding Matsubara correlation function is given as
\begin{gather}
 P_q^{\alpha_1,\dots,\alpha_q}(i\varepsilon_{n_{1}},\dots,i\varepsilon_{n_{q}}) =\! 
 \sum\limits_{\{k_1,\dots,k_q\}} \mu_{k_1,\dots, k_s} \bigl \langle  R_{k_1,\dots, k_s} \bigr \rangle ,
 \notag \\
 R_{k_1,\dots, k_s}=
 \prod\limits_{r=k_1}^{k_s}  
   \tr Q_{n_{j_1}n_{j_2}}^{\alpha_{j_1} \alpha_{j_2}} Q_{n_{j_2}n_{j_3}}^{\alpha_{j_2}\alpha_{j_3}}\dots Q_{n_{j_r}n_{j_1}}^{\alpha_{j_r}\alpha_{j_1}} .
   \label{eq:PqM}
 \end{gather}
The summation in the right hand side of Eq. \eqref{eq:PqM} is performed over all partitions~\footnote{The partitions is a set of positive integer numbers $\{k_1,\dots,k_s\}$ which satisfy the following conditions: $k_1{+}k_2{+}\dots k_s {=}q$ and $k_1{\geqslant} k_2 {\geqslant} \dots {\geqslant} k_s{>}0$.} of the integer number $q$. We note that  all replica indices in Eq. \eqref{eq:Kq} are different:
$\alpha_j {\neq} \alpha_k$ if $j{\neq} k$ for $j,k {=} 1, \dots, q$. One coefficient among the set $\{\mu_{k_1,\dots, k_s}\}$ can be chosen arbitrary. We adopt the following convention:  
$\mu_{1,1,\dots,1} {=} 1$.
We mention that the following simplified operator 
\begin{align}
K_q = & \frac{1}{4^{q}}  \sum_{p_k =\pm} \left ( \prod\limits _{j=1}^q p_j \lim\limits_{{n_{j}}\to 0}\right )
\notag \\ \times & 
P_q^{\alpha_1,\dots,\alpha_q}(ip_1|\varepsilon_{n_{1}}|,\dots,ip_q|\varepsilon_{n_{q}}|) 
 \label{eq:Kq:simp}
 \end{align}
 has the same renormalization as $\mathcal{K}_q(E_1,\dots,E_q)$. Therefore, in what follows we will work with $K_q$ instead of $\mathcal{K}_q$.

In the absence of interaction, $\gamma{=}0$, the {\NLSM} action reduces to Eq. \eqref{eq:S:int}. Then one can project $Q$ matrix to the $2{\times} 2$ subspace of a given positive and a given negative Matsubara frequencies \footnote{It is possible since the Matsubara indices of the $Q$ matrix are not mixed in the absence of interaction (the energy of diffusive modes conserves).}. The projection corresponds to reduction of ${\rm Sp}(2N)$ to ${\rm Sp}(4 N_r)$. The effective action becomes  invariant under rotations $Q {\to} \texttt{T}^{-1} Q \texttt{T}$ with $\texttt{T} {\in} {\rm U}(2N_r)$. This allows one to average operators $\mathcal{K}_q$ over ${\rm U}(2N_r)$ rotations. The resulting rotationally invariant operators can be classified with respect to the irreducible representations of ${\rm Sp}(2N)$. Each irreducible representation contains single rotationally invariant pure scaling operator \cite{Wegner1986,Gruzberg2013,Karcher2021}. The corresponding eigenoperators can uniquely be characterized by the Young tableau $\lambda{=}\textsf{(k}_\textsf{1},\dots,\textsf{k}_\textsf{s}\textsf{)}$ (with $|\lambda|{=}\sum_{\textsf{j}{=}1}^\textsf{s} \textsf{k}_\textsf{j} {=}q$). 

Although interaction breaks beautiful mathematical structure of {\NLSM} manifold, surprisingly, it does not spoil the structure of non-${\rm U}(2N_rN_m)$-invariant eigenoperators $\mathcal{K}_\lambda$ \cite{Burmistrov2013,Burmistrov2015m,Repin2016,Babkin2022}. The coefficients $\mu_{k_1,\dots, k_s}$ for $|\lambda|{=}2,3,4$ are listed in Table \ref{tab}. 
Not surprisingly, the anomalous dimension of $\mathcal{K}_\lambda$ is changed by the interaction. The renormalized eigenoperator can be written as
\begin{equation}
\mathcal{K}_\lambda = Z^q M_\lambda \mathcal{K}_\lambda[\Lambda] ,
\end{equation}
where the factor $Z$ describing renormalization of the local density of states is governed by the following RG equation
\begin{equation}
\eta_{\textsf{(1)}}= - \frac{d\ln Z}{d\ell}   = \bigl [1- 3 \ln(1+\gamma)\bigr ]t+ O(t^2) .
\label{eq:Z:rg}
\end{equation} 
We note that in the presence of interaction the expression for $\eta_{\textsf{(1)}}$ is known upto the one-loop approximation only. The quantity $M_\lambda$ determines the anomalous dimension
\begin{gather}
\eta_{\lambda} = - \frac{d\ln M_\lambda}{d\ell} = \mu_{2,1,\dots,1} t [1+3c(\gamma)t] +  O(t^3) ,
\label{eq:eta:final:q}
\end{gather}
where $\mu_{2,1,\dots,1}$ is a coefficient in expansion of the eigenoperator in series in the basis operators $R_{k_1,\dots, k_s}$, see Eq. \eqref{eq:PqM}. For the eigenoperator characterized by the Young tableau $\lambda{=}\textsf{(k}_\textsf{1},\dots,\textsf{k}_\textsf{s}\textsf{)}$ this coefficient is given as \cite{Karcher2021} (see Table \ref{tab} for $|\lambda|{=}2,3,4$) 
\begin{equation}
\mu_{2,1,\dots,1} = -\frac{1}{2} \sum_{\textsf{j}=1}^\textsf{s} \textsf{k}_\textsf{j}(\textsf{c}_\textsf{j}+2+\textsf{k}_\textsf{j}), \quad \textsf{c}_\textsf{j}=1-4 \textsf{j} .
\label{eq:eta:final:q:2}
\end{equation} 
The function $c(\gamma)$ contains information about interaction and is given as
\cite{Burmistrov2013,Burmistrov2015m,Repin2016,Babkin2022}
\begin{equation}
c(\gamma) = 2 + \frac{1+\gamma}{2\gamma}\ln^2(1+\gamma)+\frac{2+\gamma}{\gamma}\li2(-\gamma) .
\label{eq:def:c:gamma}
\end{equation}

The anomalous dimensions $\eta_\lambda$ determine the scaling with the system size $L$ of the eigen operators at the fixed point, 
\begin{equation}
\mathcal{K}_{\lambda}\sim L^{-x_{\lambda}}, \quad x_{\lambda}= |\lambda| x_\textsf{(1)}+\Delta_{\lambda} .
\label{eq:scaling:OP}
\end{equation}
Here the exponent $x_\textsf{(1)}$ coincides with the magnitude of $\eta_\textsf{(1)}$ at the fixed point, $x_\textsf{(1)}{=}\eta_\textsf{(1)}^{*}$. Similarly, the exponent $\Delta_{\lambda}$ is equal to the anomalous dimension of $M_\lambda$ at the fixed point, $\Delta_{\lambda}{=}\eta_\lambda$. 

Next we discuss how Eqs. \eqref{eq:Z:rg} and \eqref{eq:eta:final:q} are modified for the local eigenoperators situated near the boundary.

\begin{table}[t]
\caption{The coefficients $\mu_{k_1,\dots,k_s}$ for eigenoperators with $q=2,3,4$.}
\begin{tabular}{c|ccccc}
%$q=2$ & & & & & \\
${\bf q=2}$ & $\mu_{1,1}$ & $\mu_2$ & & & \\
$\textsf{(2)}$ & 1 & -1 & & &\\
$\textsf{(1,1)}$ & 1 & 2 & & &\\
\hline
%$q=3$ & & & & & \\
${\bf q=3}$ & $\mu_{1,1,1}$ & $\mu_{2,1}$ & $\mu_3$ & & \\
$\textsf{(3)}$ & 1 & -3 & 2 & & \\
$\textsf{(2,1)}$ & 1 & 1 & -2  & &\\
 $\textsf{(1,1,1)}$ & 1 & 6 & 8 & & \\
 \hline
% $q=4$ & & & & & \\ 
${\bf q=4}$ & $\mu_{1,1,1,1}$ & $\mu_{2,1,1}$ & $\mu_{3,1}$ & $\mu_{2,2}$ & $\mu_4$ \\
$\textsf{(4)}$ & 1 & -6 & 8 & 3 & -6 \\
$\textsf{(3,1)}$ & 1 & -1 & -2 & -2 & 4 \\
$\textsf{(2,2)}$ & 1 & 2 & -8 & 7 & -2 \\
$\textsf{(2,1,1)}$ & 1 & 5 & 4  & -2&-8\\
 $\textsf{(1,1,1,1)}$ & 1 & 12 & 32 & 12 & 48 
 \end{tabular}
 \label{tab}
 \end{table}

\section{Generalized surface multifractality\label{Sec:GSM:S}}

In this section we compute anomalous dimensions of the RG eigenoperators without derivatives near the boundary.

\subsection{Operator with a single $Q$ matrix}

We start analysis  from the local eigenoperator with a single $Q$ matrix, 
\begin{gather}
P^\alpha_1(i\varepsilon_n) =  \tr \langle Q_{nn}^{\alpha\alpha} \rangle.
\label{eq:P1:1}
\end{gather}
Physically, it corresponds to the average local density of states near the boundary. 
Substituting the expansion $Q{\simeq}\Lambda {+} W {-} \Lambda W^2/2$, we find that 
\begin{gather}
P^\alpha_1(i\varepsilon_n) =2  Z^{(s)} (i\varepsilon_n) \sgn \varepsilon_n ,
\label{eq:P1:2}
\end{gather}
where 
\begin{align}
Z^{(s)} (i\varepsilon_n) & =  1 -\frac{\textsf{v}}{g} \hat{\mathcal{D}}(2i|\varepsilon_n|;\bm{r},\bm{r}) 
\notag \\
& +  \frac{12\pi T\gamma}{g D} \sum_{\omega_m>|\varepsilon_n|} 
%\mathcal{D}_p(i\omega_m)
\widehat{\DD^t}(i\omega_m;\bm{r},\bm{r}) .
\label{eq:P1:2}
\end{align} 
Assuming that the point $\bm{r}$ is close to the boundary at $x=0$, we find that 
\begin{subequations}
\begin{align}
\hat{\mathcal{D}}(i\omega_n;\bm{r},\bm{r})  & \simeq 2 \int_q \mathcal{D}_q(i\omega_n) ,\label{eq:D:rel} \\
\widehat{\DD^t}(i\omega_n;\bm{r},\bm{r}) & \simeq 2 \int_q \DD^t_q(i\omega_n) .
\label{eq:DDt:rel}
\end{align}
\end{subequations}
Therefore, we find 
\begin{gather}
\mathcal{K}_{\textsf{(1)}} {=} Z^{(s)} \mathcal{K}_{\textsf{(1)}}[\Lambda], \quad  Z^{(s)} {=} 
1{+}\left (\frac{\textsf{v}}{2} {-} 3 \ln(1{+}\gamma)\right )  
 \frac{2 t h^{\epsilon}}{\epsilon} \notag \\
 {=} 1 {+}\left ({\textsf{v}}/{2} {-} 3 \ln(1{+}\gamma)\right )  
 {2 t h^{\prime \epsilon}}/{\epsilon}  .
\end{gather}
Applying the minimal subtraction scheme, we deduce the anomalous dimension of the operator $\mathcal{K}_{\textsf{(1)}}$,
\begin{equation}
\eta_{\textsf{(1)}}^{(s)} = -\frac{d\ln Z^{(s)}}{d\ell} =  2 t \left [1 - 3 \ln(1+\gamma)\right ]  + O(t^2) .
\label{eq:Z:1loop}
\end{equation}
We note that similar to the bulk anomalous dimension $\eta_{\textsf{(1)}}$ the interaction affects the anomalous dimension of $Z^{(s)}$ already in the one-loop approximation. The effect of the boundary is a factor $2$ in front of $t$ in the right hand side of Eq. \eqref{eq:Z:1loop}, cf. Eq. \eqref{eq:Z:rg}. Since the two-loop expression for the bulk anomalous dimension $\eta_{\textsf{(1)}}$ is not known at the moment, in this work we restrict our computation of the surface anomalous dimension $\eta_{\textsf{(1)}}^{(s)}$ to the one-loop order only. As we shall see below, cf. Sec. \ref{Sec:AMsurf}, one-loop renormalization of $Z^{(s)}$ will be enough in order to determine surface anomalous dimensions of eigenoperators with $q{\geqslant}2$ within two-loop approximation.

\subsection{Local eigenoperators with two $Q$ matrices}

\subsubsection{One-loop renormalization}

As known  there are two local eigenoperators with two $Q$ matrices denoted as $\mathcal{K}_{\textsf{(2)}}$ and $\mathcal{K}_{\textsf{(1,1)}}$. It will be convenient to consider the irreducible part of the corresponding correlation function,
\begin{equation}
P_2^{\alpha\beta; {\rm (irr)}}(i\varepsilon_n,i\varepsilon_m) \!=\! \llangle \tr Q_{nn}^{\alpha\alpha} \tr Q_{mm}^{\beta\beta} \rrangle + \mu_2 \langle \tr Q_{nm}^{\alpha\beta} Q_{mn}^{\beta\alpha} \rangle .
\end{equation}
Here $\mu_2{=}{-}1$ and $2$ corresponds to the operator $\mathcal{K}_{\textsf{(2)}}$ and $\mathcal{K}_{\textsf{(1,1)}}$, respectively. We note that the full correlation function can be restored as follows 
\begin{equation}
P_2^{\alpha\beta}(i\varepsilon_n,i\varepsilon_m){=} (2 Z^{(s)})^2 \sgn\varepsilon_n \sgn\varepsilon_m {+} P_2^{\alpha\beta;{\rm (irr)}}(i\varepsilon_n,i\varepsilon_m).
\end{equation}

After expansion of $Q$ to the first order in $W$,  the one-loop contribution becomes
\begin{gather}
P_{2,1}^{\alpha\beta; {\rm (irr)}}(i\varepsilon_n,i\varepsilon_m) =\mu_2 \langle \tr W_{nm}^{\alpha\beta} W_{mn}^{\beta\alpha}\rangle
=  \frac{16\mu_2}{g} \notag \\
\times \frac{1-\sgn \varepsilon_n\sgn \varepsilon_m}{2}  \hat{\mathcal{D}}(i|\varepsilon_n|+i|\varepsilon_m|;\bm{r},\bm{r}) .
\label{eq:P21:r}
\end{gather} 

Neglecting the energy dependence in the diffusive propagators and using Eq. \eqref{eq:D:rel}, we find the following one-loop result for the irreducible part of the operator $K_2$,
\begin{equation}
K_{2,1}^{\rm (irr)}= 2 t \mu_2 h^\epsilon/\epsilon .
\label{eq:K21:irr} 
\end{equation}
We note the same additional factor $2$ as in the one-loop expression for $Z^{(s)}$.

\subsubsection{Two-loop renormalization}

Next expanding $Q$ to the second order in $W$, we obtain the two-loop contribution as 
\begin{gather}
{}\hspace{-0.3cm} P_{2,2}^{\alpha\beta;{\rm (irr)}}{=}\frac{1}{4} \sgn \varepsilon_n\sgn \varepsilon_m \Llangle\tr 
(W^2)_{nn}^{\alpha\alpha}\tr (W^2)_{mm}^{\beta\beta}\Rrangle
\notag\\
{+} \mu_2 \frac{1{+}\sgn \varepsilon_n\sgn \varepsilon_m}{8} \Bigl \langle \tr (W^2)_{nm}^{\alpha\beta} (W^2)_{mn}^{\beta\alpha}\Bigr \rangle
\notag \\
{+}\mu_2 \frac{1{-}\sgn\varepsilon_n\sgn\varepsilon_m}{2} \Llangle \tr W_{nm}^{\alpha\beta} W_{mn}^{\beta\alpha} 
\begin{bmatrix}
S_{\rm 0}^{(4)}{+}S_{\rm int}^{(4)}\\
{+}\frac{1}{2} (S_{\rm int}^{(3)})^2
\end{bmatrix} 
\Rrangle .
\label{eq:K2:2:qw}
\end{gather}
In order to compute \eqref{eq:K2:2:qw}, we need to calculate several contractions of $W$ matrices. At first, using Eq. \eqref{eq:prop:full}, we find
\begin{gather}
\Llangle \tr 
(W^2)_{nn}^{\alpha\alpha}\tr (W^2)_{mm}^{\beta\beta}\Rrangle
{=} \frac{64}{g^2} \left (\hat{\mathcal{D}}(i|\varepsilon_n|{+}i|\varepsilon_m|;\bm{r},\bm{r}) \right )^2
\notag \\
\simeq \frac{64}{g^2} \left (2 \int_q \mathcal{D}_q(i|\varepsilon_n|{+}i|\varepsilon_m|) \right )^2
\to 16 \frac{(2 t)^2h^{2\epsilon}}{\epsilon^2} .
\label{eq:trw22}
\end{gather}
In the last line we use Eq. \eqref{eq:D:rel} and neglect the energy dependence in the propagators. 

Next, we proceed as follows
\begin{gather}
\Bigl \langle \tr (W^2)_{nm}^{\alpha\beta} (W^2)_{mn}^{\beta\alpha}\Bigr \rangle \hspace{5cm} {}\notag \\
{=} {-}  
3 \frac{2^7 \pi T \gamma}{g^2 D} \! \sum_{\varepsilon_k>0}  \!
\hat{\mathcal{D}}(i|\varepsilon_m|{+}i\varepsilon_k;\bm{r},\bm{r})\widehat{\DD^t}(i|\varepsilon_n|{+}i\varepsilon_k;\bm{r},\bm{r})\notag\\
 +
\frac{32\textsf{v}}{g^2} \!\hat{\mathcal{D}}(2i|\varepsilon_n|;\bm{r},\bm{r}) \hat{\mathcal{D}}(i|\varepsilon_n|{+}i|\varepsilon_m|;\bm{r},\bm{r}) 
+(n\leftrightarrow m) 
\notag \\
\simeq -  
3 \frac{2^9 \pi T \gamma}{g^2 D}  \! \sum_{\varepsilon_k>0}  \!
\int_{qp}\mathcal{D}_q(i|\varepsilon_m|{+}i\varepsilon_k)\DD^t_p(i|\varepsilon_n|{+}i\varepsilon_k)\notag\\
 +
\frac{32\textsf{v}}{g^2} 4 \int_{qp} \!\mathcal{D}_q(2i|\varepsilon_n|) \mathcal{D}_p(i|\varepsilon_n|{+}i|\varepsilon_m|) 
+(n\leftrightarrow m) 
\notag\\
%\to  16 \textsf{v} \frac{(2t)^2h^{2\epsilon}}{\epsilon^2}- 12 \frac{128\gamma}{g^2} J_{101}^0(1+\gamma) \notag \\
{\to}  16 \textsf{v} \frac{(2t)^2h^{2\epsilon}}{\epsilon^2}
{-} 48 \frac{(2t)^2h^{2\epsilon}}{\epsilon^2}
 \Bigl [ \ln(1{+}\gamma){-}\frac{\epsilon}{4}\ln^2(1{+}\gamma)\Bigr] .
\label{eq:trw23}
\end{gather}
Here we use Eqs. \eqref{eq:D:rel} and \eqref{eq:DDt:rel}. 
We refer a reader to Ref. \cite{Burmistrov2015m} for details on computation of integrals over momenta and frequency involved in Eq. \eqref{eq:trw23}.

Next we have to introduce the following non-Gaussian terms stemming from the expansion of $Q$ matrix in powers of $W$ of the {\NLSM} action:
\begin{gather}
S_{\rm 0}^{(4)}{+}S_{\rm h}^{(4)}  {=} {-} \frac{g}{64}\!\int\limits_{\bm{r}} \!\sum_{\alpha_i,n_i}
\Bigl ( 
\nabla_{12} \nabla_{34}
{+} \nabla_{14}\nabla_{23}
{+} \frac{\omega_{n_{12}+n_{34}}}{D}
\notag \\
{+}2h^2
\Bigr ) \tr \Bigl [(w(\bm{r}_1))_{n_1n_2}^{\alpha_1\alpha_2}
(w^\dag(\bm{r}_2))_{n_2n_3}^{\alpha_2\alpha_3}
\notag \\
\times
(w(\bm{r}_3))_{n_3n_4}^{\alpha_3\alpha_4}
(w^\dag(\bm{r}_4))_{n_4n_1}^{\alpha_4\alpha_1}
\Bigr ] \Biggl |_{\bm{r}_i{=}\bm{r}}
\end{gather}
(here we use a shorthand notation $\nabla_{12}{\equiv} \nabla_1{+}\nabla_2$) and
\begin{subequations}
\begin{align}
S^{(3)}_{\rm int}  & = \frac{\pi T \Gamma_t}{4}  \sum_{\alpha,n} \int\limits_{\bm{r}} \Tr I^{\alpha}_{n} \bm{\textsf{s}} W
\Tr I^{\alpha}_{-n} \bm{\textsf{s}}\Lambda W^2 ,\\
S^{(4)}_{\rm int} & = -\frac{\pi T\Gamma_t }{16} \sum_{\alpha,n} 
\int\limits_{\bm{r}} \Tr I^{\alpha}_{n} \bm{\textsf{s}} \Lambda W^2
\Tr I^{\alpha}_{-n} \bm{\textsf{s}}\Lambda W^2 .
\end{align}
\end{subequations}
Performing averaging with the help of Wick theorem and Eq. \eqref{eq:prop:full}, we obtain
\begin{widetext}
\begin{gather}
\Llangle\! \tr [W_{nm}^{\alpha\beta}W_{mn}^{\beta\alpha}] \Bigl[S_{\rm 0}^{(4)}{+}S_{\rm h}^{(4)}\Bigr] \!\Rrangle 
{=} {-} \frac{8 \textsf{v}}{g^2} \int \limits_{\bm{r}^\prime} 
\Bigl [\nabla_{12}\nabla_{34}{+}\nabla_{14}\nabla_{32} {+} 
\frac{2|\varepsilon_n|{+}|\omega_{nm}|}{D}{+}2h^2\Bigr ]
\hat{\mathcal{D}}(i2 |\varepsilon_n|;\bm{r_1},\bm{r_2})\hat{\mathcal{D}}(i|\omega_{nm}|;\bm{r_3},\bm{r})
\notag \\
\times
  \hat{\mathcal{D}}(i|\omega_{nm}|;\bm{r},\bm{r_4})\Biggl |_{\bm{r_i}{=}\bm{r}^\prime}{+} \frac{96 \pi T \gamma}{g^2 D} 
 \sum_{\omega_k>|\varepsilon_{n}|} 
 \Bigl [\nabla_{12}\nabla_{34}{+}\nabla_{14}\nabla_{32} {+} 
\frac{|\omega_k|{+}|\omega_{nm}|}{D}{+}2h^2\Bigr ]
 \widehat{\DD^t}(i|\omega_{k}|;\bm{r_1},\bm{r_2})
  \hat{\mathcal{D}}(i|\omega_{nm}|;\bm{r_3},\bm{r})
  \notag \\
\times 
  \hat{\mathcal{D}}(i|\omega_{nm}|;\bm{r},\bm{r_4}) \Biggl |_{\bm{r_i}=\bm{r}^\prime}
  {+}(n{\leftrightarrow} m) 
{=}{-} \frac{16 \textsf{v}}{g^2}
 \int_{qp} \Bigl (p^2{+}q^2{+}\frac{2|\varepsilon_n|{+}|\omega_{nm}|}{D}{+}2h^2\Bigr )
 \mathcal{D}_p(i2 |\varepsilon_n|) \mathcal{D}_{q}^2(i|\omega_{nm}|)
  {-} \frac{16 \textsf{v}}{g^2}  \int_{qp}
 \Bigl (4p_x^2 \notag \\
{+}2p_xq_x{+} p^2{+}q^2{+}\frac{2|\varepsilon_n|{+}|\omega_{nm}|}{D}{+}2h^2\Bigr )
 \mathcal{D}_p(i2 |\varepsilon_n|)
 \mathcal{D}_{q}(i|\omega_{nm}|)
 \mathcal{D}_{q_x{+}2p_x,\bm{q_{\|}}}(i|\omega_{nm}|)
 {+} 2\frac{96 \pi T \gamma}{g^2 D} 
 \sum_{\omega_k{>}|\varepsilon_{n}|} 
 \int_{qp} \Bigl (4 p_x^2{+}2p_xq_x {+}p^2
 \notag \\
 {+}q^2{+}\frac{|\omega_k|{+}|\omega_{nm}|}{D}{+}2h^2\Bigr ) \DD^t_p(i\omega_k)
 \mathcal{D}_{q}(i|\omega_{nm}|)
 \mathcal{D}_{q_x{+}2p_x,\bm{q_{\|}}}(i|\omega_{nm}|)
{+} \frac{192 \pi T \gamma}{g^2 D} 
 \sum_{\omega_k{>}|\varepsilon_{n}|} 
 \int_{qp} \Bigl (p^2{+}q^2{+}\frac{|\omega_k|{+}|\omega_{nm}|}{D}{+}2h^2\Bigr )
 \notag \\ 
 {\times}
 \DD^t_p(i\omega_k)\mathcal{D}_{\bm{q}}^2(i|\omega_{nm}|)
 {+}(n{\leftrightarrow} m) 
 {\to}  {-} 5 \textsf{v}  \frac{(2t)^2 h^{2\epsilon}}{\epsilon^2} 
  {+} 6 \frac{(2t)^2 h^{2\epsilon}}{\epsilon^2} 
 \Bigl [5 \ln(1{+}\gamma) {+}\frac{\epsilon\gamma}{1{+}\gamma} \Bigr ] 
 {+} \frac{192\gamma}{g^2}I^0_{110}(1{+}\gamma).
\label{eq:trw24}
\end{gather}
Here we introduced the following notation 
\begin{equation}
[\mathcal{D}_{q_x,\bm{q_{\|}}}(i\omega)]^{-1}=q_x^2+\bm{q_{\|}}^2+\omega/D+h^2 . 
\label{eq:spec:diff}
\end{equation}
We emphasize that appearance of such diffuson as defined in Eq. \eqref{eq:spec:diff}  is specific for the problem of boundary multifractality. 
The corresponding integrals are evaluated in Appendix \ref{App:2}. The definition of the integral $I^0_{110}$ is given in Appendix \ref{App:2}. Instead of computing the integral $I^0_{110}$ separately, it is convenient to calculate it in combination with two other similar integrals, see below.

The last contribution in Eq. \eqref{eq:K2:2:qw} can be evaluated using the following simplification which is possible due to different replica indices, $\alpha{\neq}\beta$:
\begin{equation}
S^{(4)}_{\rm int}{+}\frac{1}{2}(S^{(3)}_{\rm int})^2 {\to}  {-}  \sum_{\nu n} \int\limits_{\bm{r}, \bm{r^\prime}}
\Bigl [\delta(\bm{r}{-}\bm{r^\prime}){-}\frac{\gamma |\omega_n|}{D}\widehat{\mathcal{D}^t}(i|\omega_n|;\bm{r},\bm{r^\prime})\Bigr  ]
\frac{\pi T\Gamma_t }{4}\sum_{\textsf{j}{=}1}^3 \Tr I^{\nu}_{n} \sj \Lambda W^2(\bm{r})
\Tr I^{\nu}_{{-}n} \sj\Lambda W^2(\bm{r^\prime})
 .
\end{equation}
After tedious but straightforward calculations, we obtain 
\begin{gather}
\Llangle\! \tr W_{nm}^{\alpha\beta} W_{mn}^{\beta\alpha} \Bigl [S_{\rm int}^{(4)}
{+}\frac{(S_{\rm int}^{(3)})^2}{2} \Bigr ]\!\Rrangle {=}
\frac{96\pi T\gamma}{g^2 D} \!\!\! \int\limits_{\bm{r^{\prime}}, \bm{r^{\prime\prime}}}\!\!
\Bigl (\sum_{|\varepsilon_n|{>}\omega_k}\!\!{+}\!\!\sum_{|\varepsilon_m|{>}\omega_k}\Bigr )\Bigl [\frac{\gamma |\omega_k|}{D}\widehat{\mathcal{D}^t}(i|\omega_k|;\bm{r^{\prime}},\bm{r^{\prime\prime}}) {-}\delta(\bm{r^\prime}{-}\bm{r^{\prime\prime}})  \Bigr ] \hat{\mathcal{D}}(i|\varepsilon_n|{+}i|\varepsilon_m|;\bm{r},\bm{r^{\prime}})
\notag \\
{\times}
\hat{\mathcal{D}}(i|\varepsilon_n|{+}i|\varepsilon_m|;\bm{r},\bm{r^{\prime\prime}})
\hat{\mathcal{D}}(i|\varepsilon_n|{+}i|\varepsilon_m|{-}i\omega_k;\bm{r^{\prime}},\bm{r^{\prime\prime}})   
{\simeq}
\frac{192\pi T\gamma}{g^2 D} \!\!\int\limits_{qp}\! \Bigl (\sum_{|\varepsilon_n|{>}\omega_k}\!\!{+}\!\!\sum_{|\varepsilon_m|{>}\omega_k}\Bigr )
\Bigl [\frac{\gamma |\omega_k|}{D}\mathcal{D}^t_{\bm{p+q}}(i|\omega_k|){-}1\Bigr ] \mathcal{D}_q(i|\varepsilon_n|{+}i|\varepsilon_m|)
\notag \\
{\times}
\Bigl [ 
\mathcal{D}_q(i|\varepsilon_n|{+}i|\varepsilon_m|){+}
\mathcal{D}_{q_x{+}2p_x,\bm{q}_{\|}}(i|\varepsilon_n|{+}i|\varepsilon_m|)
\Bigr ] \mathcal{D}_p(i|\varepsilon_n|{+}i|\varepsilon_m|{-}i\omega_k)
{=}6 \frac{(2t)^2h^{2\epsilon}}{\epsilon^2}  
\Bigl [ 
\frac{2\gamma {-}(2{+}\gamma)\ln(1{+}\gamma)}{\gamma} 
{-}\epsilon \frac{(2{+}\gamma)\ln(1{+}\gamma)}{\gamma}
\notag \\
{-} \frac{\epsilon\gamma}{1{+}\gamma} 
{-}\epsilon \frac{2{+}\gamma}{\gamma} \Bigl (\li2({-}\gamma) 
{-}\frac{1}{4}\ln^2(1{+}\gamma)\Bigr )
\Bigr ] 
{-}\frac{192 \gamma}{g^2}\Bigl [I^0_{110}(1){-}\gamma I^1_{111}(1{+}\gamma)
\Bigr ].
\label{eq:trw25}
\end{gather}
Here $\li2(z){=}\sum_{k{=}1}^\infty z^k/k^2$ denotes the polylogarithm. Again we emphasize the emergence of boundary diffusons \eqref{eq:spec:diff} in the expression \eqref{eq:trw25}.
Combing the above results, Eqs. \eqref{eq:trw22}--\eqref{eq:trw24} and \eqref{eq:trw25}, we find
\begin{gather}
K_{2,2}^{\rm (irr)} {=} \Bigl \{ \mu_2(\textsf{v}{-}6\ln(1{+}\gamma)) {+} 
1{+}\frac{\mu_2 \textsf{v}}{8} {-}\frac{3\mu_2}{2} f(\gamma) 
{+}\frac{3 \epsilon \mu_2}{4}
\Bigl [\frac{2{+}3\gamma}{4\gamma} \ln^2(1+\gamma) 
{+}  \frac{2{+}\gamma}{\gamma} \Bigl (\li2({-}\gamma){+}\ln(1{+}\gamma)\Bigr )
 \Bigr ]  
\Bigr \} \frac{(2t)^2h^{2\epsilon}}{\epsilon^2}
\notag \\ 
{-}\frac{24 \gamma\mu_2}{g^2}\Bigl [I^0_{110}(1{+}\gamma){-}I^0_{110}(1){+}\gamma I^1_{111}(1{+}\gamma) \Bigr ].
\label{eq:K22:irr:0} 
\end{gather}
\end{widetext}
Using the result for the combination of $I$-integrals from Eq. \eqref{eq:app:I:res} in Appendix \ref{App:2}, we obtain 
\begin{equation}
K_{2,2}^{\rm (irr)} {=} \Bigl [ \mu_2(\textsf{v}{-}6\ln(1{+}\gamma)) {+} (b_2^{(2)}{+}\epsilon \mu_2 b_3)\Bigr ] \frac{(2t)^2h^{2\epsilon}}{\epsilon^2},  
\label{eq:K22:irr} 
\end{equation}
where
\begin{gather} 
b_2^{(2)} = 1 + \frac{\mu_2 \textsf{v}}{8} -  \frac{3 \mu_2}{2} f(\gamma) , \quad 
b_3 = \frac{3}{4} \Bigl \{\frac{2+3\gamma}{4\gamma} \ln^2(1+\gamma)
\notag \\
+  \frac{2+\gamma}{\gamma} \Bigl [\li2(-\gamma)+\ln(1+\gamma)\Bigr ]
-\frac{\gamma}{4} \Phi(\gamma)
 \Bigr \}.
\end{gather}
Here we introduced the function $\Phi(\gamma) {=} \ln^2(1{+}\gamma)/\gamma$ (see Eqs. \eqref{eq:app:Phi} and \eqref{eq:app:Phi:f}). We note that $\Phi(\gamma)$ appears from the combination of $I$-integrals. 

\subsubsection{Anomalous dimension\label{Sec:AMsurf}}

Employing the one-loop (see Eq. \eqref{eq:K21:irr}) and two-loop (see Eq. \eqref{eq:K22:irr}) results, we write the operator $K_2$ in the following form 
\begin{equation}
K_2 = (Z^{(s)})^2 M^{(s)}_2 K_2[\Lambda] .
\end{equation}
Here $K_2[\Lambda]=1$ is the classical value of $K_2$ and 
\begin{gather}
M^{(s)}_2 = 
1 +Z^{-2} (K_{2,1}^{\rm (irr)}+K_{2,2}^{\rm (irr)})
= 1 + \mu_2 \frac{2 t h^\epsilon}{\epsilon}+(b_2^{(2)}
\notag \\ +\epsilon \mu_2 b_3) 
\frac{(2t)^2 h^{2\epsilon}}{\epsilon^2}=
1 + \mu_2 \frac{2t h^{\prime \epsilon}}{\epsilon}+(b_2^{(2)}+\epsilon\mu_2  \tilde{b}_3)  \frac{(2t)^2 h^{\prime 2\epsilon}}{\epsilon^2} .
\label{eq:K2:int:1}
\end{gather}
where $\tilde{b}_3{=}b_3{+}b/4$ with $b$ given by Eq. \eqref{eq:hprime:ren}. 
Next we applying the minimal subtraction scheme to Eq. \eqref{eq:K2:int:1}. We note that the following relation holds (for $\mu_2{=}{-}1$ and $2$)
\begin{equation}
2\mu_2(2\mu_2-a_1)=8b_2^{(2)}
\end{equation}
that guarantees the finiteness of the anomalous dimension at $\epsilon\to 0$. 
 Hence, we obtain the anomalous dimensions for two eigenoperators $\mathcal{K}_{\textsf{(2)}}$ and $\mathcal{K}_{\textsf{(1,1)}}$ at the boundary 
\begin{gather}
\begin{array}{cc}
\mu_2=-1, \qquad & \eta_{\textsf{(2)}}^{(s)} = - 2t (1+6c(\gamma) t) +O(t^3) ,\\
 \mu_2=2, \qquad & \eta_{\textsf{(1,1)}}^{(s)} = 4 t (1+6c(\gamma) t)+O(t^3) .
\end{array}
\label{eq:anom:q2}
\end{gather}

\subsection{Local eigenoperators with arbitrary number of $Q$ matrices}

The above results for the local eigenoperators with two $Q$ matrix can be extended to the case of arbitrary number of $Q$ matrices in the same way as it has been done for the bulk generalized multifractality (see Ref. \cite{Babkin2022}). The eigenoperator with the number $q$ of the $Q$ matrices involved characterized by the Young tableau $\lambda{=}\textsf{(k}_\textsf{1},\dots,\textsf{k}_\textsf{s}\textsf{)}$ (with $\sum_{\textsf{j}{=}1}^\textsf{s} \textsf{k}_\textsf{j} {=}|\lambda|$), becomes
\begin{equation}
\mathcal{K}_\lambda = (Z^{(s)})^{|\lambda|} M^{(s)}_\lambda \mathcal{K}_\lambda[\Lambda] .
\end{equation}
The quantity $M^{(s)}_\lambda$ determines the anomalous dimension
\begin{gather}
\eta^{(s)}_{\lambda} = - \frac{d\ln M^{(s)}_\lambda}{d\ell} = 2 \mu_{2,1,\dots,1} t  [1+6c(\gamma)t] +  O(t^3) .
\label{eq:eta:final:q:surface}
\end{gather}
where $\mu_{2,1,\dots,1}$ is given by Eq. \eqref{eq:eta:final:q:2}. Equation \eqref{eq:eta:final:q:surface} is the main result of our work. 

The anomalous dimensions $\eta^{(s)}_\lambda$ determine the scaling with the system size $L$ of the eigen operators near the boundary at criticality, 
\begin{equation}
\mathcal{K}_{\lambda}\sim L^{-x^{(s)}_{\lambda}}, \quad x^{(s)}_{\lambda}= |\lambda| x^{(s)}_\textsf{(1)}+\Delta^{(s)}_{\lambda} .
\label{eq:scaling:OP}
\end{equation}
Here the exponent $x^{(s)}_\textsf{(1)}$ coincides with the magnitude of $\eta^{(s)}_\textsf{(1)}$, given by Eq. \eqref{eq:Z:1loop}, at the fixed point, $x^{(s)}_\textsf{(1)}{=}\eta_\textsf{(1)}^{(s)*}$. Similarly, the exponent $\Delta^{(s)}_{\lambda}$ is equal to the anomalous dimension of $M^{(s)}_\lambda$ at the fixed point, $\Delta^{(s)}_{\lambda}{=}\eta^{(s)}_\lambda$. 

\section{Discussions and conclusions\label{Sec:Final}}

\subsection{Generalization to higher orders in $t$} 

In this paper we determine the anomalous dimensions of the local eigenoperators 
situated near the boundary for the symmetry class C in the presence of interaction. We apply perturbative renormalization group expansion for the anomalous dimensions upto the second order in $t$. It was known that bulk and surface anomalous dimensions within the first order in $t$ are related by the factor $2$. We find that the same factor $2$ appears within the second order. Interestingly, that it happens in spite of the fact that the two-loop contribution to anomalous dimension is non-trivial function of interaction strength $\gamma$. 

Na\"ive idea could be that the bulk and surface exponents are related by the factor $2$ in all orders of expansion in $t$. However, it is definitely not the case for spin quantum Hall transition in $d{=}2$ dimensions. The set of bulk and surface exponents which are known exactly from mapping to percolation \cite{Gruzberg1999,Beamond2002,Mirlin2003,Evers2003,Subramaniam2008,Karcher2022} are not related by a factor of $2$, e.g. $x_\textsf{(2)}{=}1/4$ while $x^{(s)}_\textsf{(2)}{=}1/3$. Additionally, numerical computation of $x_\lambda$ and $x^{(s)}_\lambda$ indicate that the ratio between them is not a universal factor equal $2$ \cite{Babkin2023}. 

Having in mind the above discussion, it would be interesting to develop scenario how 
a factor of $2$ in weak coupling transforms into nontrivial factors different for different operators. Is the factor $2$  a feature of the perturbative expansion to all orders in $t$ while non-perturbative instanton effects are responsible for transformation to nontrivial factors? Or is the factor $2$ limited to the lowest order terms of the series in $t$ only?

\subsection{Relations to other symmetry classes}

The results reported in this paper for the surface anomalous dimensions can be directly translated to the standard Wigner-Dyson classes (classes A, AI, and AII) where the bulk generalized multifractality in the presence of interaction has been recently developed \cite{Burmistrov2013,Burmistrov2015m,Repin2016}. Similarly, within two-loop approximation the boundary multifractal exponents are twice larger than the bulk ones. Moreover, our results can be extended to the other two superconducting classes, CI and DIII, that allow interaction within the Finkel'stein {\NLSM}.

\subsection{The role of topology} 
 
Similar to the class A, the {\NLSM} for class C allows the presence of the topological $\theta$-term. The topological term does not change the classification of the local pure scaling operators but, certainly, contributes to their anomalous dimensions. At weak disorder, $t{\ll}1$, where the instanton effects can be treated in a controlled manner, the question how instantons affect the anomalous dimension of an arbitrary local operator is still not well understood. The only exception is the anomalous dimensions of bilinear in $Q$ eigenoperators for class A in the absence of interaction \cite{Pruisken2005}. Instantons are expected to affect both bulk and boundary anomalous dimensions.

\subsection{Breakdown of the Weyl symmetry}

In the absence of interaction, the Weyl-group invariance \cite{Gruzberg2013}
forces not only the bulk generalized multifractal dimensions $x_\lambda$ but also surface generalized multifractal dimensions $x^{(s)}_\lambda$  to obey the symmetry relations \cite{Babkin2023}. These symmetry relations make the exponents $x^{(s)}_\lambda$ to be the same for the eigenoperators related by the following symmetry operations: reflection, $\textsf{k}_\textsf{j}{\to} {-}\textsf{c}_\textsf{j} {-}\textsf{k}_\textsf{j}$, and permutation of some pair, $\textsf{k}_\textsf{j/i}{\to} \textsf{k}_\textsf{i/j}{+}(\textsf{c}_\textsf{i/j}{-}\textsf{c}_\textsf{j/i})/2$. Our one-loop results for the boundary anomalous dimensions are consistent with the Weyl-group invariance symmetry in the absence of interaction. The presence of interaction is known to break the symmetry relations between exponents characterizing bulk generalized multifractality \cite{Babkin2022}. Similar situation -- interaction-induced breaking of Weyl symmetry relations -- occurs with the surface exponents within two-loop approximation considered in this paper. To illustrate how it occurs let us consider the Mott-Anderson transition in $d=2+\epsilon$ dimensions. Then as follows from Eq. \eqref{eq:RG:one-loop}, there is a line of fixed points at $t_*=\epsilon/(1+6f(\gamma))$ with arbitrary $\gamma$. The surface generalized multifractal exponents becomes (to the order $\epsilon$)
\begin{equation}
x_\lambda^{(s)} =  \frac{\epsilon}{[1+6f(\gamma)]} \sum_{\textsf{j}=1}^\textsf{s} \textsf{k}_\textsf{j}(-\textsf{c}_\textsf{j}-3\ln(1+\gamma)-\textsf{k}_\textsf{j}) .
\label{eq:xLambda:2+e}
\end{equation} 
The above expression is inconsistent with Weyl symmetry in the presence of interaction, $\gamma{\neq}0$. It occurs due to appearance of $\gamma$-dependence in $x_{\textsf{(1)}}^{(s)}$. Such a situation suggests also breaking Weyl symmetry for $\gamma{\neq}0$ at the spin quantum Hall transition in $d{=}2$. Unfortunately, present numerical computing power \cite{Slevin2012,Slevin2014,Amini2014,Lee2018} is not enough to access generalized multifractal exponents and to check our predictions, in particular, to test violation of symmetry relations in the presence of interaction.

\subsection{Summary}

To summarize we developed the theory of the generalized boundary  multifractality in class C in the presence of electron-electron interaction. 
Employing two-loop renormalization group approximation controlled by inverse spin conductance $t$, 
we computed the anomalous dimensions of the pure scaling operators at the boundary of the sample. At one-loop approximation we found expected result that the boundary anomalous dimensions are two times larger than the bulk ones. Surprisingly, we found that the same relation (a factor $2$ difference) holds within two-loop approximation in spite of the nontrivial dependence of bulk and surface anomalous dimensions on interaction parameter $\gamma$. Consequently, we showed that the presence of interaction invalidates exact symmetry relations between generalized surface multifractal exponents which are consequence of Weyl symmetry in the noninteracting case. We discussed future developments and applications of our theory.

\begin{acknowledgements}
		
The authors are grateful to J. Karcher and A. Mirlin for collaboration on the related project. We thank I. Gruzberg and A. Mirlin for useful discussions and comments. The research was  supported by Russian Science Foundation (grant No. 22-42-04416).  

\end{acknowledgements}
	
\appendix

%\section{Diffuson in the presence of a boundary}
	
\begin{widetext}	
\section{Evaluation of contractions \label{App:2}}

\subsection{Eq. \eqref{eq:trw24}}

We start from rewriting the integrals over momenta in Eq. \eqref{eq:trw24} as follows
\begin{gather}
\Llangle \tr W_{nm}^{\alpha\beta} W_{mn}^{\beta\alpha} \Bigl[S_{\rm 0}^{(4)}+S_{\rm h}^{(4)}\Bigr ]\Rrangle \to 
-  \frac{32 \textsf{v}}{g^2}
\int_{qp} \Bigl [ \mathcal{D}_p(0) \mathcal{D}_{q}(0) + \mathcal{D}^2_{q}(0)\Bigr ]
+ \frac{384 \pi T \gamma}{g^2 D} \int_{qp} \sum_{\omega>0} \Bigl [ 
\DD^t_p(i\omega)
 \mathcal{D}_{q}(0)+ \mathcal{D}^t_p(i\omega)
 \mathcal{D}_{q}^2(0)
\Bigr ] \notag \\
-  \frac{32 \textsf{v}}{g^2}
\int_{qp} \Bigl [\mathcal{D}_{p_x-q_x,\bm{q_{\|}}}(0)
\mathcal{D}_{p_x+q_x,\bm{q_{\|}}}(0)
+ \mathcal{D}_{p}(0)
\mathcal{D}_{p_x+q_x,\bm{q_{\|}}}(0)+2 p_x^2 \mathcal{D}_{p}(0)
\mathcal{D}_{p_x+q_x,\bm{q_{\|}}}(0) \mathcal{D}_{p_x-q_x,\bm{q_{\|}}}(0)\Bigr ]
\notag \\
+ \frac{384 \pi T \gamma}{g^2 D} \int_{qp} \sum_{\omega>0} \Bigl [  
\mathcal{D}^t_p(i\omega)
 \mathcal{D}_{q}(0)
 + \DD^t_p(i\omega)+
 2p_x(q_x+2p_x) \DD^t_p(i\omega)\mathcal{D}_{q}(0)
 \Bigr ]
 \mathcal{D}_{q_x+2p_x,\bm{q_{\|}}}(0) .
\end{gather}
Next we find
\begin{gather}
\Llangle \tr W_{nm}^{\alpha\beta} W_{mn}^{\beta\alpha} \Bigl[S_{\rm 0}^{(4)}+S_{\rm h}^{(4)}\Bigr ]\Rrangle \to 
- 4 \textsf{v}  \frac{(2t)^2h^{2\epsilon}}{\epsilon^2} - \frac{64  \textsf{v}}{g^2}I_1
+\frac{192\gamma}{g^2} \Bigl [2 J_{110}^0(1+\gamma)+J_{020}^0(1+\gamma)
+I^0_{110}(1+\gamma) \notag \\ +2 \ln(1+\gamma) I_1\Bigr ]
\to - 4 \textsf{v} \left (1+\frac{1}{4}\right )\frac{(2t)^2h^{2\epsilon}}{\epsilon^2} + 24 \frac{(2t)^2h^{2\epsilon}}{\epsilon^2} \Bigl [ \Bigr (1+\frac{1}{4}\Bigl ) \ln(1+\gamma) + \frac{\epsilon\gamma}{4(1+\gamma)}\Bigr ]+ \frac{192\gamma}{g^2}I^0_{110}(1+\gamma)
\end{gather}

Here we introduce the following notations for integral over momenta and frequency,
\begin{gather}
J^\delta_{\nu\mu\eta}(a)= \int_{qp} \int\limits_0^\infty ds \, s^\delta \frac{1}{(p^2+h^2+s)^\nu}\frac{1}{(p^2+h^2+as)} \frac{1}{(q^2+h^2)^\mu}\frac{1}{(\bm{p}+\bm{q})^2+h^2+s)^\eta} 
\end{gather}
Also we used the following relations
\begin{gather}
\int_{q} \mathcal{D}_{q}(0) = - \frac{2 \Omega_d h^\epsilon \Gamma(1-\epsilon/2)\Gamma(1+\epsilon/2)}{\epsilon}  , \qquad \int_{qp} \mathcal{D}^2_{q}(0) = \int_{qp}
\mathcal{D}_{p_x-q_x,\bm{q_{\|}}}(0)
\mathcal{D}_{p_x+q_x,\bm{q_{\|}}}(0) = 0 ,\notag \\
\frac{2\pi T}{D} \sum_{\omega>0} 
\DD^t_p(i\omega) =\frac{\ln(1+\gamma)}{\gamma}\mathcal{D}_{p}(0) , \\
\frac{2\pi T\gamma}{D} \int_{qp} \sum_{\omega>0} 
 2p_x(q_x{+}2p_x) \DD^t_p(i\omega)\mathcal{D}_{q}(0)
 \mathcal{D}_{q_x{+}2p_x,\bm{q_{\|}}}(0)
 = \frac{2\pi T\gamma}{D} \int_{qp} \sum_{\omega>0} 
 2p_x(q_x{+}p_x) \DD^t_p(i\omega)\mathcal{D}_{q_x{-}p_x,\bm{q_{\|}}}(0)
 \mathcal{D}_{q_x{+}p_x,\bm{q_{\|}}}(0) 
 \notag \\
 = 2 \ln(1+\gamma) \int_{qp} p_x^2 \mathcal{D}_{p}(0)\mathcal{D}_{q_x-p_x,\bm{q_{\|}}}(0)
 \mathcal{D}_{q_x+p_x,\bm{q_{\|}}}(0) = 2 \ln(1+\gamma) I_1 ,\notag
 %\frac{A_\epsilon h^{2\epsilon}}{\epsilon^2}
\end{gather}
where $\Omega_d=S_d/[2(2\pi)^d]$ and $S_d=2\pi^{d/2}/\Gamma(d/2)$ is the area of the $d$-dimensional sphere. We use $t = 4\Omega_d/g$ at arbitrary dimensionality  such that  $t= 1/(\pi g)$ at $d=2$. The integral $I_1$ is evaluated as follows

\begin{gather}
I_1=\int_{qp} p_x^2 \mathcal{D}_{p}(0)
\mathcal{D}_{p_x+q_x,\bm{q_{\|}}}(0) \mathcal{D}_{p_x-q_x,\bm{q_{\|}}}(0) = 
\int_0^1 dz \int_{qp} \frac{p_x^2}{(p^2+h^2)(\bm{q_{\|}}^2+(q_x+p_x(1-2 z))^2+4p_x^2 z(1-z)+h^2)^2} \notag\\
= \int_0^1 dz \int_{qp} \frac{p_x^2}{(p^2+h^2)(4p_x^2 z(1-z)+h^2)^{2-d/2}(q^2+h^2)^2} = h^{d-4}\Omega_d\frac{\Gamma(d/2)\Gamma(2-d/2)}{\Gamma(2)}\frac{\Gamma(3-d/2)}{\Gamma(2-d/2)} 
 \int_0^1  dz \int_0^1 du \notag \\
 \times
\int_{p} \frac{p_x^2 u^{1-d/2}}{(\bm{p_{\|}}^2(1-u)+p_x^2 [1-u(1-2z)^2]+h^2)^{3-d/2}}
= h^{2\epsilon}\Omega^2_d\frac{\Gamma(d/2)\Gamma(2-d/2)}{\Gamma(2)}\frac{\Gamma(3-d/2)}{\Gamma(2-d/2)} 
\frac{\Gamma(d/2+1)\Gamma(2-d)}{d \Gamma(3-d/2)} \notag \\
\times \int_0^1  du \int_0^1 dv
u^{1-d/2} (1-u)^{-(d-1)/2} (1-u v^2)^{-3/2} =-\frac{h^{2\epsilon}\Omega^2_d}{2\epsilon} \Gamma^2(d/2)\Gamma(3-d)\int_0^1  du\, u^{1-d/2} (1-u)^{-d/2}
\notag \\
= -\frac{h^{2\epsilon}\Omega^2_d}{2\epsilon} \Gamma^2(d/2)\Gamma(3-d) 
\frac{\Gamma(2-d/2)\Gamma(1-d/2)}{\Gamma(3-d)}
=\frac{A_\epsilon h^{2\epsilon}}{\epsilon^2} ,
\end{gather}
where $A_\epsilon= \Omega_d^2 \Gamma^2(1-\epsilon/2)\Gamma^2(1+\epsilon/2)$.
The evaluation of integrals $J^\delta_{\nu\mu\eta}(a)$ is described in Ref.  \cite{Burmistrov2015m}. Also we introduced the following new integrals
\begin{gather}
I^\delta_{\nu\mu\eta}(a)= \int_{qp} \int\limits_0^\infty ds \, s^\delta \frac{1}{(q^2+h^2)^\nu}\frac{1}{(p^2+h^2+as)} \frac{1}{((q_x+2p_x)^2+\bm{q_{\|}}^2+h^2)^\mu}\frac{1}{(\bm{p}+\bm{q})^2+h^2+s)^\eta} .
\end{gather}

\subsection{Eq. \eqref{eq:trw25}}

\begin{gather}
\Llangle \tr W_{nm}^{\alpha\beta} W_{mn}^{\beta\alpha} \Bigl [S_{\rm int}^{(4)}
+\frac{1}{2} (S_{\rm int}^{(3)})^2\Bigr ]\Rrangle \to - \frac{384\pi T\gamma}{g^2 D} \int_{qp} \sum_{\omega>0}
 \Bigl [1-\frac{\gamma \omega}{D}\mathcal{D}^t_{\bm{p+q}}(i\omega)\Bigr ] \mathcal{D}_p(i\omega)
\Bigl [ \mathcal{D}_q(0) +\mathcal{D}_{q_x+2p_x,\bm{q}_{\|}}(0)\Bigr ]\notag \\
-\frac{192 \gamma}{g^2}\Bigl [ J^{0}_{020}(1)-\gamma J^{1}_{021}(1+\gamma)
+I^0_{110}(1)-\gamma I^1_{111}(1+\gamma)
\Bigr ] = - 6 \frac{(2t)^2h^{2\epsilon}}{\epsilon^2}  
\Bigl [ 
-\frac{2\gamma -(2+\gamma)\ln(1+\gamma)}{\gamma} 
+ \frac{\epsilon\gamma}{1+\gamma} 
\notag \\
+\epsilon \frac{(2+\gamma)\ln(1+\gamma)}{\gamma}
+\epsilon \frac{2+\gamma}{\gamma} \Bigl (\li2(-\gamma) 
+\frac{1}{4}\ln^2(1+\gamma)\Bigr )
\Bigr ] -\frac{192 \gamma}{g^2}\Bigl [I^0_{110}(1)-\gamma I^1_{111}(1+\gamma)
\Bigr ]
\end{gather}
Here we used the known results for the integrals $J^\delta_{\nu\mu\eta}(a)$ from Ref.  \cite{Burmistrov2015m}. Instead of computation of integrals $I^0_{110}$, $I^0_{110}$, and $I^1_{111}$ separately, it is more convenient to evaluate the combination they appear together:
 \begin{gather}
 I^0_{110}(1+\gamma)-I^0_{110}(1)+\gamma I^1_{111}(1+\gamma) = 
 \gamma \int_{qp} \int\limits_0^\infty ds \, s \Bigl [\frac{1}{(\bm{p}+\bm{q})^2+(1+\gamma)s+h^2)}-\frac{1}{(p^2+(1+\gamma)s+h^2)} \Bigr ] 
 \frac{1}{(p^2+s+h^2)}\notag \\
 \times \frac{1}{(q^2+h^2)((q_x+2p_x)^2+\bm{q_{\|}}^2+h^2)}
 = - \gamma \int_{qp} \int\limits_0^\infty ds 
 \frac{(q^2+2\bm{p q}) s}{((\bm{p}+\bm{q})^2+(1+\gamma)s+h^2)(p^2+(1+\gamma)s+h^2)(p^2+s+h^2)}\notag \\
 \times \frac{1}{(q^2+h^2)((q_x+2p_x)^2+\bm{q_{\|}}^2+h^2)} =
 \int_{qp} \int\limits_0^\infty ds \Bigl [ 
 \frac{1}{(p^2+(1+\gamma)s+h^2)}-\frac{1}{(p^2+s+h^2)}\Bigr] \frac{(q^2+2\bm{p q})}{(q^2+h^2)((q_x+2p_x)^2+\bm{q_{\|}}^2+h^2)}\notag \\
 \times \frac{1}{((\bm{p}+\bm{q})^2+(1+\gamma)s+h^2)}
 = - \int_{qp} \int\limits_0^\infty ds 
\frac{(q^2+2\bm{p q})}{(p^2+s+h^2)((\bm{p}+\bm{q})^2+(1+\gamma)s+h^2)(q^2+h^2)((q_x+2p_x)^2+\bm{q_{\|}}^2+h^2)} .
 \end{gather}
Here in the last line we employed  the following transformation $\bm{p}\to\bm{P}+\bm{Q}$ and $\bm{q}\to-\bm{Q}$ that makes $q^2+2\bm{p q}\to-Q^2-2\bm{P Q}$.  
Now we employ the Feynman trick and find
\begin{gather}
 I^0_{110}(1{+}\gamma){-}I^0_{110}(1){+}\gamma I^1_{111}(1+\gamma) {=} 
 {-} \Gamma(4) \int_0^1dz \int dx_1 dx_2 dx_3 \delta(1{-}x_1{-}x_2{-}x_3) \int_{qp} \int\limits_0^\infty ds
  (q^2{+}2\bm{p q}) \Bigl [x_1(\bm{p_{\|}}{+}z \bm{q_{\|}})^2 \notag \\
  {+} (x_2{+}x_3{+}z(1{-}z)x_1)\bm{q_{\|}}^2 {+}(x_1{+}4x_3)\Bigl (p_x{+} \frac{zx_1{+}2x_3}{x_1{+}4x_3}q_x\Bigr )^2{+} \frac{(x_1(x_2{+}x_3{+}z(1{-}z)x_1){+}4x_2x_3)}{x_1{+}4x_3} q_x^2{+}
 (1{+}\gamma z)s {+}h^2\Bigr ]^{-4}
 \end{gather}
Performing integration over momenta and frequency and using the parametrization $x_1=s/(s+1)$, $x_2=u/(s+1)$, $x_3=(1-u)/(s+1)$, where $0\leqslant s<\infty$ and $0\leqslant u\leqslant 1$ (with the Jacobian $1/(s+1)^3$), we obtain
 \begin{gather}
 I^0_{110}(1{+}\gamma){-}I^0_{110}(1){+}\gamma I^1_{111}(1+\gamma) {=}
 \frac{h^{2\epsilon}\Omega_d^2}{2\epsilon} \Gamma^2(d/2)\Gamma(3{-}d)
  \int_0^1dz \frac{(1{-}2z)}{1{+}\gamma z} \int_0^\infty ds \int_0^1 du (s{+}1)^{d{-}2} s^{(1{-}d)/2}
 \notag \\
 \times  
  [1{+}z(1{-}z) s]^{(1{-}d)/2} \Bigl [ s(1{+}z(1{-}z) s){+}4 u(1{-}u)\Bigr]^{{-}1/2}\Bigl [ \frac{d{-}1}{1{+}z(1{-}z) s}+\frac{s}{s(1{+}z(1{-}z) s){+}4 u(1{-}u)}
 \Bigr ] .
 \end{gather}
 The integrals over $z$, $s$, and $u$ are convergent in $d=2$, therefore, we can set $d=2$. Then we find
 \begin{gather}
 I^0_{110}(1{+}\gamma){-}I^0_{110}(1){+}\gamma I^1_{111}(1+\gamma) {=}
\frac{h^{2\epsilon}A_\epsilon}{2\epsilon} \int_0^1dz \frac{(1{-}2z)}{1{+}\gamma z} \int_0^\infty ds \frac{1}{\sqrt{s(1{+}z(1{-}z) s)^3}}\Bigl [ \arctan \frac{1}{\sqrt{s(1{+}z(1{-}z) s)}}\notag \\
 {+} \frac{\sqrt{s(1{+}z(1{-}z) s)}}{[1+s(1{+}z(1{-}z) s)]} \Bigr]{=} \frac{h^{2\epsilon}A_\epsilon}{2\epsilon}  
 \int_0^1dz \frac{(1{-}2z)}{1{+}\gamma z}  \int_0^\infty dy \Bigl [
 \frac{1}{\sqrt{z(1-z)}}\frac{1}{\sqrt{y(1{+}y)^3}}\arctan\frac{\sqrt{z(1-z)}}{\sqrt{y(1{+}y)}}\notag \\
  {+} \frac{1}{(y{+}1)[z(1{-}z){+}y(1{+}y)]}\Bigr ] = \frac{h^{2\epsilon}A_\epsilon}{2\epsilon}  
 \int_0^1 \frac{dz}{1{+}\gamma z} \Bigl [ \frac{(1{-}2z)}{\sqrt{z(1{-}z)}} \int_0^\infty \frac{dv}{\cosh^2(v/2)} 
 \arctan\frac{2\sqrt{z(1{-}z)}}{\sinh v} + \frac{\ln(1{-}z)}{z}-\frac{\ln z}{1{-}z} \Bigr ]
 \notag \\
 =\frac{h^{2\epsilon}A_\epsilon}{2\epsilon}   \Phi(\gamma) .
 \label{eq:app:I:res}
\end{gather}
Here we introduced $y=z(1-z) s$ and $v = 2 \arcsinh\sqrt{y}$. The function $\Phi(\gamma)$ is given as follows
\begin{gather}
\Phi(\gamma)=\int_0^1dz \frac{F(z)}{1{+}\gamma z} , 
\qquad F(z)= -(1{-}2z) \Bigl (\frac{\ln z}{1{-}z} +
\frac{\ln(1-z)}{z}\Bigr )
 {+} \frac{\ln(1{-}z)}{z}-\frac{\ln z}{1{-}z} {=} 2\ln(1{-}z) {-}2 \ln z  .
 \label{eq:app:Phi}
\end{gather}
Finally, integrating over $z$ exactly, we find
\begin{gather}
\Phi(\gamma) {=} \frac{\ln^2(1{+}\gamma)}{\gamma}
\label{eq:app:Phi:f}
\end{gather}
 \end{widetext}

\bibliography{literature_classC}	

\begin{thebibliography}{75}%
\makeatletter
\providecommand \@ifxundefined [1]{%
 \@ifx{#1\undefined}
}%
\providecommand \@ifnum [1]{%
 \ifnum #1\expandafter \@firstoftwo
 \else \expandafter \@secondoftwo
 \fi
}%
\providecommand \@ifx [1]{%
 \ifx #1\expandafter \@firstoftwo
 \else \expandafter \@secondoftwo
 \fi
}%
\providecommand \natexlab [1]{#1}%
\providecommand \enquote  [1]{``#1''}%
\providecommand \bibnamefont  [1]{#1}%
\providecommand \bibfnamefont [1]{#1}%
\providecommand \citenamefont [1]{#1}%
\providecommand \href@noop [0]{\@secondoftwo}%
\providecommand \href [0]{\begingroup \@sanitize@url \@href}%
\providecommand \@href[1]{\@@startlink{#1}\@@href}%
\providecommand \@@href[1]{\endgroup#1\@@endlink}%
\providecommand \@sanitize@url [0]{\catcode `\\12\catcode `\$12\catcode
  `\&12\catcode `\#12\catcode `\^12\catcode `\_12\catcode `\%12\relax}%
\providecommand \@@startlink[1]{}%
\providecommand \@@endlink[0]{}%
\providecommand \url  [0]{\begingroup\@sanitize@url \@url }%
\providecommand \@url [1]{\endgroup\@href {#1}{\urlprefix }}%
\providecommand \urlprefix  [0]{URL }%
\providecommand \Eprint [0]{\href }%
\providecommand \doibase [0]{http://dx.doi.org/}%
\providecommand \selectlanguage [0]{\@gobble}%
\providecommand \bibinfo  [0]{\@secondoftwo}%
\providecommand \bibfield  [0]{\@secondoftwo}%
\providecommand \translation [1]{[#1]}%
\providecommand \BibitemOpen [0]{}%
\providecommand \bibitemStop [0]{}%
\providecommand \bibitemNoStop [0]{.\EOS\space}%
\providecommand \EOS [0]{\spacefactor3000\relax}%
\providecommand \BibitemShut  [1]{\csname bibitem#1\endcsname}%
\let\auto@bib@innerbib\@empty
%</preamble>
\bibitem [{\citenamefont {Anderson}(1958)}]{Anderson58}%
  \BibitemOpen
  \bibfield  {author} {\bibinfo {author} {\bibfnamefont {P.~W.}\ \bibnamefont
  {Anderson}},\ }\bibfield  {title} {\enquote {\bibinfo {title} {Absence of
  diffusion in certain random lattices},}\ }\href
  {http://link.aps.org/doi/10.1103/PhysRev.109.1492} {\bibfield  {journal}
  {\bibinfo  {journal} {Phys. Rev.}\ }\textbf {\bibinfo {volume} {109}},\
  \bibinfo {pages} {1492} (\bibinfo {year} {1958})}\BibitemShut {NoStop}%
\bibitem [{\citenamefont {Wegner}(1980)}]{Wegner1980}%
  \BibitemOpen
  \bibfield  {author} {\bibinfo {author} {\bibfnamefont {F.}~\bibnamefont
  {Wegner}},\ }\bibfield  {title} {\enquote {\bibinfo {title} {Inverse
  participation ratio in $2+\epsilon$ dimensions},}\ }\href
  {http://link.springer.com/article/10.1007/BF01325284} {\bibfield  {journal}
  {\bibinfo  {journal} {Z. Phys.B}\ }\textbf {\bibinfo {volume} {36}},\
  \bibinfo {pages} {209} (\bibinfo {year} {1980})}\BibitemShut {NoStop}%
\bibitem [{\citenamefont {Castellani}\ and\ \citenamefont
  {Peliti}(1986)}]{Castellani1986}%
  \BibitemOpen
  \bibfield  {author} {\bibinfo {author} {\bibfnamefont {C.}~\bibnamefont
  {Castellani}}\ and\ \bibinfo {author} {\bibfnamefont {L.}~\bibnamefont
  {Peliti}},\ }\bibfield  {title} {\enquote {\bibinfo {title} {Multifractal
  wavefunction at the localisation threshold},}\ }\href
  {http://stacks.iop.org/0305-4470/19/i=8/a=004} {\bibfield  {journal}
  {\bibinfo  {journal} {J. Phys. A}\ }\textbf {\bibinfo {volume} {19}},\
  \bibinfo {pages} {L429} (\bibinfo {year} {1986})}\BibitemShut {NoStop}%
\bibitem [{\citenamefont {Mirlin}(2000)}]{Mirlin2000}%
  \BibitemOpen
  \bibfield  {author} {\bibinfo {author} {\bibfnamefont {A.~D.}\ \bibnamefont
  {Mirlin}},\ }\bibfield  {title} {\enquote {\bibinfo {title} {Statistics of
  energy levels and eigenfunctions in disordered systems},}\ }\href
  {http://www.sciencedirect.com/science/article/pii/S0370157399000915}
  {\bibfield  {journal} {\bibinfo  {journal} {Phys. Rep.}\ }\textbf {\bibinfo
  {volume} {326}},\ \bibinfo {pages} {259} (\bibinfo {year}
  {2000})}\BibitemShut {NoStop}%
\bibitem [{\citenamefont {Evers}\ and\ \citenamefont
  {Mirlin}(2008)}]{EversMirlin}%
  \BibitemOpen
  \bibfield  {author} {\bibinfo {author} {\bibfnamefont {F.}~\bibnamefont
  {Evers}}\ and\ \bibinfo {author} {\bibfnamefont {A.~D.}\ \bibnamefont
  {Mirlin}},\ }\bibfield  {title} {\enquote {\bibinfo {title} {Anderson
  transitions},}\ }\href {http://link.aps.org/doi/10.1103/RevModPhys.80.1355}
  {\bibfield  {journal} {\bibinfo  {journal} {Rev. Mod. Phys.}\ }\textbf
  {\bibinfo {volume} {80}},\ \bibinfo {pages} {1355} (\bibinfo {year}
  {2008})}\BibitemShut {NoStop}%
\bibitem [{\citenamefont {H{$\mathrm{\ddot{o}}$}f}\ and\ \citenamefont
  {Wegner}(1986)}]{Wegner1986}%
  \BibitemOpen
  \bibfield  {author} {\bibinfo {author} {\bibfnamefont {D.}~\bibnamefont
  {H{$\mathrm{\ddot{o}}$}f}}\ and\ \bibinfo {author} {\bibfnamefont
  {F.}~\bibnamefont {Wegner}},\ }\bibfield  {title} {\enquote {\bibinfo {title}
  {Calculation of anomalous dimensions for the nonlinear sigma model},}\ }\href
  {http://www.sciencedirect.com/science/article/pii/0550321386905754}
  {\bibfield  {journal} {\bibinfo  {journal} {Nucl. Phys. B}\ }\textbf
  {\bibinfo {volume} {275}},\ \bibinfo {pages} {561} (\bibinfo {year}
  {1986})}\BibitemShut {NoStop}%
\bibitem [{\citenamefont {Gruzberg}\ \emph {et~al.}(2013)\citenamefont
  {Gruzberg}, \citenamefont {Mirlin},\ and\ \citenamefont
  {Zirnbauer}}]{Gruzberg2013}%
  \BibitemOpen
  \bibfield  {author} {\bibinfo {author} {\bibfnamefont {I.~A.}\ \bibnamefont
  {Gruzberg}}, \bibinfo {author} {\bibfnamefont {A.~D.}\ \bibnamefont
  {Mirlin}}, \ and\ \bibinfo {author} {\bibfnamefont {M.~R.}\ \bibnamefont
  {Zirnbauer}},\ }\bibfield  {title} {\enquote {\bibinfo {title}
  {Classification and symmetry properties of scaling dimensions at {Anderson}
  transitions},}\ }\href {\doibase 10.1103/PhysRevB.87.125144} {\bibfield
  {journal} {\bibinfo  {journal} {Phys. Rev. B}\ }\textbf {\bibinfo {volume}
  {87}},\ \bibinfo {pages} {125144} (\bibinfo {year} {2013})}\BibitemShut
  {NoStop}%
\bibitem [{\citenamefont {Karcher}\ \emph
  {et~al.}(2022{\natexlab{a}})\citenamefont {Karcher}, \citenamefont
  {Gruzberg},\ and\ \citenamefont {Mirlin}}]{Karcher2022}%
  \BibitemOpen
  \bibfield  {author} {\bibinfo {author} {\bibfnamefont {Jonas~F.}\
  \bibnamefont {Karcher}}, \bibinfo {author} {\bibfnamefont {Ilya~A.}\
  \bibnamefont {Gruzberg}}, \ and\ \bibinfo {author} {\bibfnamefont
  {Alexander~D.}\ \bibnamefont {Mirlin}},\ }\bibfield  {title} {\enquote
  {\bibinfo {title} {Generalized multifractality at the spin quantum {Hall}
  transition: {Percolation} mapping and pure-scaling observables},}\ }\href
  {\doibase 10.1103/PhysRevB.105.184205} {\bibfield  {journal} {\bibinfo
  {journal} {Phys. Rev. B}\ }\textbf {\bibinfo {volume} {105}},\ \bibinfo
  {pages} {184205} (\bibinfo {year} {2022}{\natexlab{a}})}\BibitemShut
  {NoStop}%
\bibitem [{\citenamefont {Karcher}\ \emph
  {et~al.}(2022{\natexlab{b}})\citenamefont {Karcher}, \citenamefont
  {Gruzberg},\ and\ \citenamefont {Mirlin}}]{Karcher2022b}%
  \BibitemOpen
  \bibfield  {author} {\bibinfo {author} {\bibfnamefont {J.~F.}\ \bibnamefont
  {Karcher}}, \bibinfo {author} {\bibfnamefont {Ilya~A.}\ \bibnamefont
  {Gruzberg}}, \ and\ \bibinfo {author} {\bibfnamefont {Alexander~D.}\
  \bibnamefont {Mirlin}},\ }\bibfield  {title} {\enquote {\bibinfo {title}
  {Generalized multifractality at metal-insulator transitions and in metallic
  phases of two-dimensional disordered systems},}\ }\href {\doibase
  10.1103/PhysRevB.106.104202} {\bibfield  {journal} {\bibinfo  {journal}
  {Physical Review B}\ }\textbf {\bibinfo {volume} {106}},\ \bibinfo {pages}
  {104202} (\bibinfo {year} {2022}{\natexlab{b}})}\BibitemShut {NoStop}%
\bibitem [{\citenamefont {Karcher}\ \emph
  {et~al.}(2023{\natexlab{a}})\citenamefont {Karcher}, \citenamefont
  {Gruzberg},\ and\ \citenamefont {Mirlin}}]{Karcher2023}%
  \BibitemOpen
  \bibfield  {author} {\bibinfo {author} {\bibfnamefont {Jonas~F.}\
  \bibnamefont {Karcher}}, \bibinfo {author} {\bibfnamefont {Ilya~A.}\
  \bibnamefont {Gruzberg}}, \ and\ \bibinfo {author} {\bibfnamefont
  {Alexander~D.}\ \bibnamefont {Mirlin}},\ }\bibfield  {title} {\enquote
  {\bibinfo {title} {Generalized multifractality in two-dimensional disordered
  systems of chiral symmetry classes},}\ }\href {\doibase
  10.1103/PhysRevB.107.104202} {\bibfield  {journal} {\bibinfo  {journal}
  {Physical Review B}\ }\textbf {\bibinfo {volume} {107}},\ \bibinfo {pages}
  {104202} (\bibinfo {year} {2023}{\natexlab{a}})}\BibitemShut {NoStop}%
\bibitem [{\citenamefont {Mirlin}\ \emph {et~al.}(2006)\citenamefont {Mirlin},
  \citenamefont {Fyodorov}, \citenamefont {Mildenberger},\ and\ \citenamefont
  {Evers}}]{Mirlin2006}%
  \BibitemOpen
  \bibfield  {author} {\bibinfo {author} {\bibfnamefont {A.~D.}\ \bibnamefont
  {Mirlin}}, \bibinfo {author} {\bibfnamefont {Y.~V.}\ \bibnamefont
  {Fyodorov}}, \bibinfo {author} {\bibfnamefont {A.}~\bibnamefont
  {Mildenberger}}, \ and\ \bibinfo {author} {\bibfnamefont {F.}~\bibnamefont
  {Evers}},\ }\bibfield  {title} {\enquote {\bibinfo {title} {Exact relations
  between multifractal exponents at the {Anderson} transition},}\ }\href
  {\doibase 10.1103/PhysRevLett.97.046803} {\bibfield  {journal} {\bibinfo
  {journal} {Phys. Rev. Lett.}\ }\textbf {\bibinfo {volume} {97}},\ \bibinfo
  {pages} {046803} (\bibinfo {year} {2006})}\BibitemShut {NoStop}%
\bibitem [{\citenamefont {Gruzberg}\ \emph {et~al.}(2011)\citenamefont
  {Gruzberg}, \citenamefont {Ludwig}, \citenamefont {Mirlin},\ and\
  \citenamefont {Zirnbauer}}]{Gruzberg2011}%
  \BibitemOpen
  \bibfield  {author} {\bibinfo {author} {\bibfnamefont {I.~A.}\ \bibnamefont
  {Gruzberg}}, \bibinfo {author} {\bibfnamefont {A.~W.~W.}\ \bibnamefont
  {Ludwig}}, \bibinfo {author} {\bibfnamefont {A.~D.}\ \bibnamefont {Mirlin}},
  \ and\ \bibinfo {author} {\bibfnamefont {M.~R.}\ \bibnamefont {Zirnbauer}},\
  }\bibfield  {title} {\enquote {\bibinfo {title} {Symmetries of multifractal
  spectra and field theories of {Anderson} localization},}\ }\href {\doibase
  10.1103/PhysRevLett.107.086403} {\bibfield  {journal} {\bibinfo  {journal}
  {Phys. Rev. Lett.}\ }\textbf {\bibinfo {volume} {107}},\ \bibinfo {pages}
  {086403} (\bibinfo {year} {2011})}\BibitemShut {NoStop}%
\bibitem [{\citenamefont {Subramaniam}\ \emph {et~al.}(2006)\citenamefont
  {Subramaniam}, \citenamefont {Gruzberg}, \citenamefont {Ludwig},
  \citenamefont {Evers}, \citenamefont {Mildenberger},\ and\ \citenamefont
  {Mirlin}}]{Subramaniam2006}%
  \BibitemOpen
  \bibfield  {author} {\bibinfo {author} {\bibfnamefont {A.~R.}\ \bibnamefont
  {Subramaniam}}, \bibinfo {author} {\bibfnamefont {I.~A.}\ \bibnamefont
  {Gruzberg}}, \bibinfo {author} {\bibfnamefont {A.~W.~W.}\ \bibnamefont
  {Ludwig}}, \bibinfo {author} {\bibfnamefont {F.}~\bibnamefont {Evers}},
  \bibinfo {author} {\bibfnamefont {A.}~\bibnamefont {Mildenberger}}, \ and\
  \bibinfo {author} {\bibfnamefont {A.~D.}\ \bibnamefont {Mirlin}},\ }\bibfield
   {title} {\enquote {\bibinfo {title} {Surface criticality and multifractality
  at localization transitions},}\ }\href {\doibase
  10.1103/PhysRevLett.96.126802} {\bibfield  {journal} {\bibinfo  {journal}
  {Physical Review Letters}\ }\textbf {\bibinfo {volume} {96}},\ \bibinfo
  {pages} {126802} (\bibinfo {year} {2006})}\BibitemShut {NoStop}%
\bibitem [{\citenamefont {Mildenberger}\ \emph {et~al.}(2007)\citenamefont
  {Mildenberger}, \citenamefont {Subramaniam}, \citenamefont {Narayanan},
  \citenamefont {Evers}, \citenamefont {Gruzberg},\ and\ \citenamefont
  {Mirlin}}]{mildenberger2007}%
  \BibitemOpen
  \bibfield  {author} {\bibinfo {author} {\bibfnamefont {A.}~\bibnamefont
  {Mildenberger}}, \bibinfo {author} {\bibfnamefont {A.~R.}\ \bibnamefont
  {Subramaniam}}, \bibinfo {author} {\bibfnamefont {R.}~\bibnamefont
  {Narayanan}}, \bibinfo {author} {\bibfnamefont {F.}~\bibnamefont {Evers}},
  \bibinfo {author} {\bibfnamefont {I.~A.}\ \bibnamefont {Gruzberg}}, \ and\
  \bibinfo {author} {\bibfnamefont {A.~D.}\ \bibnamefont {Mirlin}},\ }\bibfield
   {title} {\enquote {\bibinfo {title} {Boundary multifractality in critical
  one-dimensional systems with long-range hopping},}\ }\href {\doibase
  10.1103/PhysRevB.75.094204} {\bibfield  {journal} {\bibinfo  {journal}
  {Physical Review B}\ }\textbf {\bibinfo {volume} {75}},\ \bibinfo {pages}
  {094204} (\bibinfo {year} {2007})}\BibitemShut {NoStop}%
\bibitem [{\citenamefont {Subramaniam}\ \emph {et~al.}(2008)\citenamefont
  {Subramaniam}, \citenamefont {Gruzberg},\ and\ \citenamefont
  {Ludwig}}]{Subramaniam2008}%
  \BibitemOpen
  \bibfield  {author} {\bibinfo {author} {\bibfnamefont {A.~R.}\ \bibnamefont
  {Subramaniam}}, \bibinfo {author} {\bibfnamefont {I.~A.}\ \bibnamefont
  {Gruzberg}}, \ and\ \bibinfo {author} {\bibfnamefont {A.~W.~W.}\ \bibnamefont
  {Ludwig}},\ }\bibfield  {title} {\enquote {\bibinfo {title} {Boundary
  criticality and multifractality at the two-dimensional spin quantum {Hall}
  transition},}\ }\href {\doibase 10.1103/PhysRevB.78.245105} {\bibfield
  {journal} {\bibinfo  {journal} {Phys. Rev. B}\ }\textbf {\bibinfo {volume}
  {78}},\ \bibinfo {pages} {245105} (\bibinfo {year} {2008})}\BibitemShut
  {NoStop}%
\bibitem [{\citenamefont {Evers}\ \emph {et~al.}(2008)\citenamefont {Evers},
  \citenamefont {Mildenberger},\ and\ \citenamefont {Mirlin}}]{Evers2008}%
  \BibitemOpen
  \bibfield  {author} {\bibinfo {author} {\bibfnamefont {F.}~\bibnamefont
  {Evers}}, \bibinfo {author} {\bibfnamefont {A.}~\bibnamefont {Mildenberger}},
  \ and\ \bibinfo {author} {\bibfnamefont {A.~D.}\ \bibnamefont {Mirlin}},\
  }\bibfield  {title} {\enquote {\bibinfo {title} {Multifractality at the
  quantum {Hall} transition: {Beyond} the parabolic paradigm},}\ }\href
  {\doibase 10.1103/PhysRevLett.101.116803} {\bibfield  {journal} {\bibinfo
  {journal} {Phys. Rev. Lett.}\ }\textbf {\bibinfo {volume} {101}},\ \bibinfo
  {pages} {116803} (\bibinfo {year} {2008})}\BibitemShut {NoStop}%
\bibitem [{\citenamefont {Obuse}\ \emph {et~al.}(2008)\citenamefont {Obuse},
  \citenamefont {Subramaniam}, \citenamefont {Furusaki}, \citenamefont
  {Gruzberg},\ and\ \citenamefont {Ludwig}}]{Obuse2008}%
  \BibitemOpen
  \bibfield  {author} {\bibinfo {author} {\bibfnamefont {H.}~\bibnamefont
  {Obuse}}, \bibinfo {author} {\bibfnamefont {A.~R.}\ \bibnamefont
  {Subramaniam}}, \bibinfo {author} {\bibfnamefont {A.}~\bibnamefont
  {Furusaki}}, \bibinfo {author} {\bibfnamefont {I.~A.}\ \bibnamefont
  {Gruzberg}}, \ and\ \bibinfo {author} {\bibfnamefont {A.~W.~W.}\ \bibnamefont
  {Ludwig}},\ }\bibfield  {title} {\enquote {\bibinfo {title} {Boundary
  multifractality at the integer quantum {Hall} plateau transition:
  {Implications} for the critical theory},}\ }\href {\doibase
  10.1103/PhysRevLett.101.116802} {\bibfield  {journal} {\bibinfo  {journal}
  {Phys. Rev. Lett.}\ }\textbf {\bibinfo {volume} {101}},\ \bibinfo {pages}
  {116802} (\bibinfo {year} {2008})}\BibitemShut {NoStop}%
\bibitem [{\citenamefont {Babkin}\ \emph {et~al.}()\citenamefont {Babkin},
  \citenamefont {Karcher}, \citenamefont {Burmistrov},\ and\ \citenamefont
  {Mirlin}}]{Babkin2023}%
  \BibitemOpen
  \bibfield  {author} {\bibinfo {author} {\bibfnamefont {S.~S.}\ \bibnamefont
  {Babkin}}, \bibinfo {author} {\bibfnamefont {J.~F.}\ \bibnamefont {Karcher}},
  \bibinfo {author} {\bibfnamefont {I.~S.}\ \bibnamefont {Burmistrov}}, \ and\
  \bibinfo {author} {\bibfnamefont {A.~D.}\ \bibnamefont {Mirlin}},\
  }\href@noop {} {\enquote {\bibinfo {title} {Generalized surface
  multifractality in {2D} disordered systems},}\ }\bibinfo {howpublished}
  {arXiv:2306.09455}\BibitemShut {NoStop}%
\bibitem [{\citenamefont {Karcher}\ \emph {et~al.}(2021)\citenamefont
  {Karcher}, \citenamefont {Charles}, \citenamefont {Gruzberg},\ and\
  \citenamefont {Mirlin}}]{Karcher2021}%
  \BibitemOpen
  \bibfield  {author} {\bibinfo {author} {\bibfnamefont {Jonas~F.}\
  \bibnamefont {Karcher}}, \bibinfo {author} {\bibfnamefont {Noah}\
  \bibnamefont {Charles}}, \bibinfo {author} {\bibfnamefont {Ilya~A.}\
  \bibnamefont {Gruzberg}}, \ and\ \bibinfo {author} {\bibfnamefont
  {Alexander~D.}\ \bibnamefont {Mirlin}},\ }\bibfield  {title} {\enquote
  {\bibinfo {title} {Generalized multifractality at spin quantum {Hall}
  transition},}\ }\href {\doibase https://doi.org/10.1016/j.aop.2021.168584}
  {\bibfield  {journal} {\bibinfo  {journal} {Ann. Phys. (N.Y.)}\ }\textbf
  {\bibinfo {volume} {435}},\ \bibinfo {pages} {168584} (\bibinfo {year}
  {2021})},\ \bibinfo {note} {special issue on Philip W. Anderson}\BibitemShut
  {NoStop}%
\bibitem [{\citenamefont {Karcher}\ \emph
  {et~al.}(2023{\natexlab{b}})\citenamefont {Karcher}, \citenamefont
  {Gruzberg},\ and\ \citenamefont {Mirlin}}]{Karcher2023a}%
  \BibitemOpen
  \bibfield  {author} {\bibinfo {author} {\bibfnamefont {Jonas~F.}\
  \bibnamefont {Karcher}}, \bibinfo {author} {\bibfnamefont {Ilya~A.}\
  \bibnamefont {Gruzberg}}, \ and\ \bibinfo {author} {\bibfnamefont
  {Alexander~D.}\ \bibnamefont {Mirlin}},\ }\bibfield  {title} {\enquote
  {\bibinfo {title} {Metal-insulator transition in a two-dimensional system of
  chiral unitary class},}\ }\href {\doibase 10.1103/PhysRevB.107.L020201}
  {\bibfield  {journal} {\bibinfo  {journal} {Physical Review B}\ }\textbf
  {\bibinfo {volume} {107}},\ \bibinfo {pages} {L020201} (\bibinfo {year}
  {2023}{\natexlab{b}})}\BibitemShut {NoStop}%
\bibitem [{\citenamefont {Mascheck}\ \emph {et~al.}(2012)\citenamefont
  {Mascheck}, \citenamefont {Schmidt}, \citenamefont {Silies}, \citenamefont
  {Yatsui}, \citenamefont {Kitamura}, \citenamefont {Ohtsu}, \citenamefont
  {Leipold}, \citenamefont {Runge},\ and\ \citenamefont
  {Lienau}}]{Mascheck2012}%
  \BibitemOpen
  \bibfield  {author} {\bibinfo {author} {\bibfnamefont {M.}~\bibnamefont
  {Mascheck}}, \bibinfo {author} {\bibfnamefont {S.}~\bibnamefont {Schmidt}},
  \bibinfo {author} {\bibfnamefont {M.}~\bibnamefont {Silies}}, \bibinfo
  {author} {\bibfnamefont {T.}~\bibnamefont {Yatsui}}, \bibinfo {author}
  {\bibfnamefont {K.}~\bibnamefont {Kitamura}}, \bibinfo {author}
  {\bibfnamefont {M.}~\bibnamefont {Ohtsu}}, \bibinfo {author} {\bibfnamefont
  {D.}~\bibnamefont {Leipold}}, \bibinfo {author} {\bibfnamefont
  {E.}~\bibnamefont {Runge}}, \ and\ \bibinfo {author} {\bibfnamefont
  {C.}~\bibnamefont {Lienau}},\ }\bibfield  {title} {\enquote {\bibinfo {title}
  {Observing the localization of light in space and time by ultrafast
  second-harmonic microscopy},}\ }\href
  {http://www.nature.com/nphoton/journal/v6/n5/full/nphoton.2012.69.html}
  {\bibfield  {journal} {\bibinfo  {journal} {Nat. Photonics}\ }\textbf
  {\bibinfo {volume} {6}},\ \bibinfo {pages} {293} (\bibinfo {year}
  {2012})}\BibitemShut {NoStop}%
\bibitem [{\citenamefont {Faez}\ \emph {et~al.}(2009)\citenamefont {Faez},
  \citenamefont {Strybulevych}, \citenamefont {Page}, \citenamefont
  {Lagendijk},\ and\ \citenamefont {van Tiggelen}}]{Faez2009}%
  \BibitemOpen
  \bibfield  {author} {\bibinfo {author} {\bibfnamefont {S.}~\bibnamefont
  {Faez}}, \bibinfo {author} {\bibfnamefont {A.}~\bibnamefont {Strybulevych}},
  \bibinfo {author} {\bibfnamefont {J.~H.}\ \bibnamefont {Page}}, \bibinfo
  {author} {\bibfnamefont {A.}~\bibnamefont {Lagendijk}}, \ and\ \bibinfo
  {author} {\bibfnamefont {B.~A.}\ \bibnamefont {van Tiggelen}},\ }\bibfield
  {title} {\enquote {\bibinfo {title} {Observation of multifractality in
  {Anderson} localization of ultrasound},}\ }\href
  {http://link.aps.org/doi/10.1103/PhysRevLett.103.155703} {\bibfield
  {journal} {\bibinfo  {journal} {Phys. Rev. Lett.}\ }\textbf {\bibinfo
  {volume} {103}},\ \bibinfo {pages} {155703} (\bibinfo {year}
  {2009})}\BibitemShut {NoStop}%
\bibitem [{\citenamefont {Richardella}\ \emph {et~al.}(2010)\citenamefont
  {Richardella}, \citenamefont {Roushan}, \citenamefont {Mack}, \citenamefont
  {Zhou}, \citenamefont {Huse}, \citenamefont {Awshalom},\ and\ \citenamefont
  {Yazdani}}]{Richardella}%
  \BibitemOpen
  \bibfield  {author} {\bibinfo {author} {\bibfnamefont {A.}~\bibnamefont
  {Richardella}}, \bibinfo {author} {\bibfnamefont {P.}~\bibnamefont
  {Roushan}}, \bibinfo {author} {\bibfnamefont {S.}~\bibnamefont {Mack}},
  \bibinfo {author} {\bibfnamefont {B.}~\bibnamefont {Zhou}}, \bibinfo {author}
  {\bibfnamefont {D.~A.}\ \bibnamefont {Huse}}, \bibinfo {author}
  {\bibfnamefont {D.~D.}\ \bibnamefont {Awshalom}}, \ and\ \bibinfo {author}
  {\bibfnamefont {A.}~\bibnamefont {Yazdani}},\ }\bibfield  {title} {\enquote
  {\bibinfo {title} {Visualizing critical correlations near the metal-insulator
  transition in {Ga$_{1-x}$Mn$_x$As}},}\ }\href
  {http://science.sciencemag.org/content/327/5966/665.full} {\bibfield
  {journal} {\bibinfo  {journal} {Science}\ }\textbf {\bibinfo {volume}
  {327}},\ \bibinfo {pages} {665} (\bibinfo {year} {2010})}\BibitemShut
  {NoStop}%
\bibitem [{\citenamefont {Liz\'ee}\ \emph {et~al.}(2023)\citenamefont
  {Liz\'ee}, \citenamefont {Stosiek}, \citenamefont {Burmistrov}, \citenamefont
  {Cren},\ and\ \citenamefont {Brun}}]{Lizee2023}%
  \BibitemOpen
  \bibfield  {author} {\bibinfo {author} {\bibfnamefont {Mathieu}\ \bibnamefont
  {Liz\'ee}}, \bibinfo {author} {\bibfnamefont {Matthias}\ \bibnamefont
  {Stosiek}}, \bibinfo {author} {\bibfnamefont {Igor}\ \bibnamefont
  {Burmistrov}}, \bibinfo {author} {\bibfnamefont {Tristan}\ \bibnamefont
  {Cren}}, \ and\ \bibinfo {author} {\bibfnamefont {Christophe}\ \bibnamefont
  {Brun}},\ }\bibfield  {title} {\enquote {\bibinfo {title} {Local density of
  states fluctuations in a two-dimensional superconductor as a probe of quantum
  diffusion},}\ }\href {\doibase 10.1103/PhysRevB.107.174508} {\bibfield
  {journal} {\bibinfo  {journal} {Physical Review B}\ }\textbf {\bibinfo
  {volume} {107}},\ \bibinfo {pages} {174508} (\bibinfo {year}
  {2023})}\BibitemShut {NoStop}%
\bibitem [{\citenamefont {Feigel'man}\ \emph {et~al.}(2007)\citenamefont
  {Feigel'man}, \citenamefont {Ioffe}, \citenamefont {Kravtsov},\ and\
  \citenamefont {Yuzbashyan}}]{FeigelmanYuzbashyan2007}%
  \BibitemOpen
  \bibfield  {author} {\bibinfo {author} {\bibfnamefont {M.~V.}\ \bibnamefont
  {Feigel'man}}, \bibinfo {author} {\bibfnamefont {L.~B.}\ \bibnamefont
  {Ioffe}}, \bibinfo {author} {\bibfnamefont {V.~E.}\ \bibnamefont {Kravtsov}},
  \ and\ \bibinfo {author} {\bibfnamefont {E.~A.}\ \bibnamefont {Yuzbashyan}},\
  }\bibfield  {title} {\enquote {\bibinfo {title} {Eigenfunction fractality and
  pseudogap state near the superconductor-insulator transition},}\ }\href
  {http://link.aps.org/doi/10.1103/PhysRevLett.98.027001} {\bibfield  {journal}
  {\bibinfo  {journal} {Phys. Rev. Lett.}\ }\textbf {\bibinfo {volume} {98}},\
  \bibinfo {pages} {027001} (\bibinfo {year} {2007})}\BibitemShut {NoStop}%
\bibitem [{\citenamefont {Feigel'man}\ \emph {et~al.}(2010)\citenamefont
  {Feigel'man}, \citenamefont {Ioffe}, \citenamefont {Kravtsov},\ and\
  \citenamefont {Cuevas}}]{FeigelmanCuevas2010}%
  \BibitemOpen
  \bibfield  {author} {\bibinfo {author} {\bibfnamefont {M.V.}\ \bibnamefont
  {Feigel'man}}, \bibinfo {author} {\bibfnamefont {L.B.}\ \bibnamefont
  {Ioffe}}, \bibinfo {author} {\bibfnamefont {V.E.}\ \bibnamefont {Kravtsov}},
  \ and\ \bibinfo {author} {\bibfnamefont {E.}~\bibnamefont {Cuevas}},\
  }\bibfield  {title} {\enquote {\bibinfo {title} {Fractal superconductivity
  near localization threshold},}\ }\href
  {http://www.sciencedirect.com/science/article/pii/S000349161000062X}
  {\bibfield  {journal} {\bibinfo  {journal} {Ann. Phys. (N.Y.)}\ }\textbf
  {\bibinfo {volume} {325}},\ \bibinfo {pages} {1390} (\bibinfo {year}
  {2010})}\BibitemShut {NoStop}%
\bibitem [{\citenamefont {Burmistrov}\ \emph {et~al.}(2012)\citenamefont
  {Burmistrov}, \citenamefont {Gornyi},\ and\ \citenamefont
  {Mirlin}}]{Burmistrov2012}%
  \BibitemOpen
  \bibfield  {author} {\bibinfo {author} {\bibfnamefont {I.~S.}\ \bibnamefont
  {Burmistrov}}, \bibinfo {author} {\bibfnamefont {I.~V.}\ \bibnamefont
  {Gornyi}}, \ and\ \bibinfo {author} {\bibfnamefont {A.~D.}\ \bibnamefont
  {Mirlin}},\ }\bibfield  {title} {\enquote {\bibinfo {title} {Enhancement of
  the critical temperature of superconductors by {Anderson} localization},}\
  }\href {\doibase 10.1103/PhysRevLett.108.017002} {\bibfield  {journal}
  {\bibinfo  {journal} {Physical Review Letters}\ }\textbf {\bibinfo {volume}
  {108}},\ \bibinfo {pages} {017002} (\bibinfo {year} {2012})}\BibitemShut
  {NoStop}%
\bibitem [{\citenamefont {Burmistrov}\ \emph
  {et~al.}(2015{\natexlab{a}})\citenamefont {Burmistrov}, \citenamefont
  {Gornyi},\ and\ \citenamefont {Mirlin}}]{Burmistrov2015b}%
  \BibitemOpen
  \bibfield  {author} {\bibinfo {author} {\bibfnamefont {I.~S.}\ \bibnamefont
  {Burmistrov}}, \bibinfo {author} {\bibfnamefont {I.~V.}\ \bibnamefont
  {Gornyi}}, \ and\ \bibinfo {author} {\bibfnamefont {A.~D.}\ \bibnamefont
  {Mirlin}},\ }\bibfield  {title} {\enquote {\bibinfo {title}
  {Superconductor-insulator transitions: phase diagram and
  magnetoresistance},}\ }\href
  {http://journals.aps.org/prb/abstract/10.1103/PhysRevB.92.014506} {\bibfield
  {journal} {\bibinfo  {journal} {Phys. Rev. B}\ }\textbf {\bibinfo {volume}
  {92}},\ \bibinfo {pages} {014506} (\bibinfo {year}
  {2015}{\natexlab{a}})}\BibitemShut {NoStop}%
\bibitem [{\citenamefont {Dell'Anna}(2013)}]{DellAnna}%
  \BibitemOpen
  \bibfield  {author} {\bibinfo {author} {\bibfnamefont {L.}~\bibnamefont
  {Dell'Anna}},\ }\bibfield  {title} {\enquote {\bibinfo {title} {Enhancement
  of critical temperatures in disordered bipartite lattices},}\ }\href
  {http://link.aps.org/doi/10.1103/PhysRevB.88.195139} {\bibfield  {journal}
  {\bibinfo  {journal} {Phys. Rev. B}\ }\textbf {\bibinfo {volume} {88}},\
  \bibinfo {pages} {195139} (\bibinfo {year} {2013})}\BibitemShut {NoStop}%
\bibitem [{\citenamefont {Burmistrov}\ \emph {et~al.}(2021)\citenamefont
  {Burmistrov}, \citenamefont {Gornyi},\ and\ \citenamefont
  {Mirlin}}]{Burmistrov2021}%
  \BibitemOpen
  \bibfield  {author} {\bibinfo {author} {\bibfnamefont {I.~S.}\ \bibnamefont
  {Burmistrov}}, \bibinfo {author} {\bibfnamefont {I.~V.}\ \bibnamefont
  {Gornyi}}, \ and\ \bibinfo {author} {\bibfnamefont {A.~D.}\ \bibnamefont
  {Mirlin}},\ }\bibfield  {title} {\enquote {\bibinfo {title}
  {Multifractally-enhanced superconductivity in thin films},}\ }\href {\doibase
  https://doi.org/10.1016/j.aop.2021.168499} {\bibfield  {journal} {\bibinfo
  {journal} {Ann. Phys. (N.Y.)}\ }\textbf {\bibinfo {volume} {435}},\ \bibinfo
  {pages} {168499} (\bibinfo {year} {2021})},\ \bibinfo {note} {special Issue
  on Localisation 2020}\BibitemShut {NoStop}%
\bibitem [{\citenamefont {Andriyakhina}\ and\ \citenamefont
  {Burmistrov}(2022)}]{Andriyakhina2022}%
  \BibitemOpen
  \bibfield  {author} {\bibinfo {author} {\bibfnamefont {E.S.}\ \bibnamefont
  {Andriyakhina}}\ and\ \bibinfo {author} {\bibfnamefont {I.S.}\ \bibnamefont
  {Burmistrov}},\ }\bibfield  {title} {\enquote {\bibinfo {title}
  {Multifractally-enhanced superconductivity in two-dimensional systems with
  spin-orbit coupling},}\ }\href@noop {} {\bibfield  {journal} {\bibinfo
  {journal} {ZhETF}\ }\textbf {\bibinfo {volume} {162}},\ \bibinfo {pages}
  {522} (\bibinfo {year} {2022})}\BibitemShut {NoStop}%
\bibitem [{\citenamefont {Nosov}\ \emph {et~al.}(2023)\citenamefont {Nosov},
  \citenamefont {Burmistrov},\ and\ \citenamefont {Raghu}}]{Nosov2023}%
  \BibitemOpen
  \bibfield  {author} {\bibinfo {author} {\bibfnamefont {P.~A.}\ \bibnamefont
  {Nosov}}, \bibinfo {author} {\bibfnamefont {I.~S.}\ \bibnamefont
  {Burmistrov}}, \ and\ \bibinfo {author} {\bibfnamefont {S.}~\bibnamefont
  {Raghu}},\ }\bibfield  {title} {\enquote {\bibinfo {title} {Interplay of
  superconductivity and localization near a two-dimensional ferromagnetic
  quantum critical point},}\ }\href {\doibase 10.1103/PhysRevB.107.144508}
  {\bibfield  {journal} {\bibinfo  {journal} {Physical Review B}\ }\textbf
  {\bibinfo {volume} {107}},\ \bibinfo {pages} {144508} (\bibinfo {year}
  {2023})}\BibitemShut {NoStop}%
\bibitem [{\citenamefont {Fan}\ and\ \citenamefont
  {Garc\'ia-Garc\'ia}(2020)}]{Fan2020}%
  \BibitemOpen
  \bibfield  {author} {\bibinfo {author} {\bibfnamefont {Bo}~\bibnamefont
  {Fan}}\ and\ \bibinfo {author} {\bibfnamefont {A.~M.}\ \bibnamefont
  {Garc\'ia-Garc\'ia}},\ }\bibfield  {title} {\enquote {\bibinfo {title}
  {Enhanced phase-coherent multifractal two-dimensional superconductivity},}\
  }\href {\doibase 10.1103/PhysRevB.101.104509} {\bibfield  {journal} {\bibinfo
   {journal} {Phys. Rev. B}\ }\textbf {\bibinfo {volume} {101}},\ \bibinfo
  {pages} {104509} (\bibinfo {year} {2020})}\BibitemShut {NoStop}%
\bibitem [{\citenamefont {Stosiek}\ \emph {et~al.}(2020)\citenamefont
  {Stosiek}, \citenamefont {Lang},\ and\ \citenamefont {Evers}}]{Stosiek2020}%
  \BibitemOpen
  \bibfield  {author} {\bibinfo {author} {\bibfnamefont {M.}~\bibnamefont
  {Stosiek}}, \bibinfo {author} {\bibfnamefont {B.}~\bibnamefont {Lang}}, \
  and\ \bibinfo {author} {\bibfnamefont {F.}~\bibnamefont {Evers}},\ }\bibfield
   {title} {\enquote {\bibinfo {title} {{Self-consistent-field ensembles of
  disordered Hamiltonians: Efficient solver and application to superconducting
  films}},}\ }\href {\doibase 10.1103/PhysRevB.101.144503} {\bibfield
  {journal} {\bibinfo  {journal} {Phys. Rev. B}\ }\textbf {\bibinfo {volume}
  {101}},\ \bibinfo {pages} {144503} (\bibinfo {year} {2020})}\BibitemShut
  {NoStop}%
\bibitem [{\citenamefont {Stosiek}\ \emph {et~al.}(2021)\citenamefont
  {Stosiek}, \citenamefont {Evers},\ and\ \citenamefont
  {Burmistrov}}]{Stosiek2021}%
  \BibitemOpen
  \bibfield  {author} {\bibinfo {author} {\bibfnamefont {M.}~\bibnamefont
  {Stosiek}}, \bibinfo {author} {\bibfnamefont {F.}~\bibnamefont {Evers}}, \
  and\ \bibinfo {author} {\bibfnamefont {I.~S.}\ \bibnamefont {Burmistrov}},\
  }\bibfield  {title} {\enquote {\bibinfo {title} {{Multifractal correlations
  of the local density of states in dirty superconducting films}},}\ }\href
  {\doibase 10.1103/PhysRevResearch.3.L042016} {\bibfield  {journal} {\bibinfo
  {journal} {Phys. Rev. Research}\ }\textbf {\bibinfo {volume} {3}},\ \bibinfo
  {pages} {L042016} (\bibinfo {year} {2021})}\BibitemShut {NoStop}%
\bibitem [{\citenamefont {Foster}\ and\ \citenamefont
  {Yuzbashyan}(2012)}]{Foster2012}%
  \BibitemOpen
  \bibfield  {author} {\bibinfo {author} {\bibfnamefont {M.~S.}\ \bibnamefont
  {Foster}}\ and\ \bibinfo {author} {\bibfnamefont {E.~A.}\ \bibnamefont
  {Yuzbashyan}},\ }\bibfield  {title} {\enquote {\bibinfo {title}
  {Interaction-mediated surface-state instability in disordered
  three-dimensional topological superconductors with spin {SU(2)} symmetry},}\
  }\href {\doibase 10.1103/PhysRevLett.109.246801} {\bibfield  {journal}
  {\bibinfo  {journal} {Phys. Rev. Lett.}\ }\textbf {\bibinfo {volume} {109}},\
  \bibinfo {pages} {246801} (\bibinfo {year} {2012})}\BibitemShut {NoStop}%
\bibitem [{\citenamefont {Foster}\ \emph {et~al.}(2014)\citenamefont {Foster},
  \citenamefont {Xie},\ and\ \citenamefont {Chou}}]{Foster2014}%
  \BibitemOpen
  \bibfield  {author} {\bibinfo {author} {\bibfnamefont {M.~S.}\ \bibnamefont
  {Foster}}, \bibinfo {author} {\bibfnamefont {H.-Y.}\ \bibnamefont {Xie}}, \
  and\ \bibinfo {author} {\bibfnamefont {Y.-Z.}\ \bibnamefont {Chou}},\
  }\bibfield  {title} {\enquote {\bibinfo {title} {Topological protection,
  disorder, and interactions: {Survival} at the surface of three-dimensional
  topological superconductors},}\ }\href
  {http://link.aps.org/doi/10.1103/PhysRevB.89.155140} {\bibfield  {journal}
  {\bibinfo  {journal} {Phys. Rev. B}\ }\textbf {\bibinfo {volume} {89}},\
  \bibinfo {pages} {155140} (\bibinfo {year} {2014})}\BibitemShut {NoStop}%
\bibitem [{\citenamefont {Kettemann}\ and\ \citenamefont
  {Mucciolo}(2006)}]{Kettemann2006}%
  \BibitemOpen
  \bibfield  {author} {\bibinfo {author} {\bibfnamefont {S.}~\bibnamefont
  {Kettemann}}\ and\ \bibinfo {author} {\bibfnamefont {E.~R.}\ \bibnamefont
  {Mucciolo}},\ }\bibfield  {title} {\enquote {\bibinfo {title} {Free magnetic
  moments in disordered systems},}\ }\href
  {http://jetpletters.ac.ru/ps/1060/article_16089.shtml} {\bibfield  {journal}
  {\bibinfo  {journal} {JETP Lett.}\ }\textbf {\bibinfo {volume} {83}},\
  \bibinfo {pages} {284} (\bibinfo {year} {2006})}\BibitemShut {NoStop}%
\bibitem [{\citenamefont {Micklitz}\ \emph {et~al.}(2006)\citenamefont
  {Micklitz}, \citenamefont {Altland}, \citenamefont {Costi},\ and\
  \citenamefont {Rosch}}]{Micklitz2006}%
  \BibitemOpen
  \bibfield  {author} {\bibinfo {author} {\bibfnamefont {T.}~\bibnamefont
  {Micklitz}}, \bibinfo {author} {\bibfnamefont {A.}~\bibnamefont {Altland}},
  \bibinfo {author} {\bibfnamefont {T.~A.}\ \bibnamefont {Costi}}, \ and\
  \bibinfo {author} {\bibfnamefont {A.}~\bibnamefont {Rosch}},\ }\bibfield
  {title} {\enquote {\bibinfo {title} {Universal dephasing rate due to diluted
  {Kondo} impurities},}\ }\href
  {http://link.aps.org/doi/10.1103/PhysRevLett.96.226601} {\bibfield  {journal}
  {\bibinfo  {journal} {Phys. Rev. Lett.}\ }\textbf {\bibinfo {volume} {96}},\
  \bibinfo {pages} {226601} (\bibinfo {year} {2006})}\BibitemShut {NoStop}%
\bibitem [{\citenamefont {Kettemann}\ and\ \citenamefont
  {Mucciolo}(2007)}]{Kettemann2007}%
  \BibitemOpen
  \bibfield  {author} {\bibinfo {author} {\bibfnamefont {S.}~\bibnamefont
  {Kettemann}}\ and\ \bibinfo {author} {\bibfnamefont {E.~R.}\ \bibnamefont
  {Mucciolo}},\ }\bibfield  {title} {\enquote {\bibinfo {title}
  {Disorder-quenched {Kondo} effect in mesoscopic electronic systems},}\ }\href
  {http://link.aps.org/doi/10.1103/PhysRevB.75.184407} {\bibfield  {journal}
  {\bibinfo  {journal} {Phys. Rev. B}\ }\textbf {\bibinfo {volume} {75}},\
  \bibinfo {pages} {184407} (\bibinfo {year} {2007})}\BibitemShut {NoStop}%
\bibitem [{\citenamefont {Feigel'man}\ and\ \citenamefont
  {Kravtsov}(2019)}]{Feigelman2019}%
  \BibitemOpen
  \bibfield  {author} {\bibinfo {author} {\bibfnamefont {M.~V.}\ \bibnamefont
  {Feigel'man}}\ and\ \bibinfo {author} {\bibfnamefont {V.~E.}\ \bibnamefont
  {Kravtsov}},\ }\bibfield  {title} {\enquote {\bibinfo {title}
  {Electron-phonon cooling power in {Anderson} insulators},}\ }\href {\doibase
  10.1103/PhysRevB.99.125415} {\bibfield  {journal} {\bibinfo  {journal} {Phys.
  Rev. B}\ }\textbf {\bibinfo {volume} {99}},\ \bibinfo {pages} {125415}
  (\bibinfo {year} {2019})}\BibitemShut {NoStop}%
\bibitem [{\citenamefont {Kettemann}(2016)}]{Kettemann2016}%
  \BibitemOpen
  \bibfield  {author} {\bibinfo {author} {\bibfnamefont {S.}~\bibnamefont
  {Kettemann}},\ }\bibfield  {title} {\enquote {\bibinfo {title} {Exponential
  orthogonality catastrophe at the {Anderson} metal-insulator transition},}\
  }\href {\doibase 10.1103/PhysRevLett.117.146602} {\bibfield  {journal}
  {\bibinfo  {journal} {Phys. Rev. Lett.}\ }\textbf {\bibinfo {volume} {117}},\
  \bibinfo {pages} {146602} (\bibinfo {year} {2016})}\BibitemShut {NoStop}%
\bibitem [{\citenamefont {Burmistrov}\ and\ \citenamefont
  {Skvortsov}(2018)}]{Burmistrov2018}%
  \BibitemOpen
  \bibfield  {author} {\bibinfo {author} {\bibfnamefont {I.~S.}\ \bibnamefont
  {Burmistrov}}\ and\ \bibinfo {author} {\bibfnamefont {M.~A.}\ \bibnamefont
  {Skvortsov}},\ }\bibfield  {title} {\enquote {\bibinfo {title} {Magnetic
  disorder in superconductors: {Enhancement} by mesoscopic fluctuations},}\
  }\href {\doibase 10.1103/PhysRevB.97.014515} {\bibfield  {journal} {\bibinfo
  {journal} {Phys. Rev. B}\ }\textbf {\bibinfo {volume} {97}},\ \bibinfo
  {pages} {014515} (\bibinfo {year} {2018})}\BibitemShut {NoStop}%
\bibitem [{\citenamefont {Babkin}\ \emph {et~al.}(2022)\citenamefont {Babkin},
  \citenamefont {Lyublinskaya},\ and\ \citenamefont
  {Burmistrov}}]{Babkin2022Y}%
  \BibitemOpen
  \bibfield  {author} {\bibinfo {author} {\bibfnamefont {S.~S.}\ \bibnamefont
  {Babkin}}, \bibinfo {author} {\bibfnamefont {A.~A.}\ \bibnamefont
  {Lyublinskaya}}, \ and\ \bibinfo {author} {\bibfnamefont {I.~S.}\
  \bibnamefont {Burmistrov}},\ }\bibfield  {title} {\enquote {\bibinfo {title}
  {Broadened {Yu-Shiba-Rusinov} states in dirty superconducting films and
  heterostructures},}\ }\href {\doibase 10.1103/PhysRevResearch.4.023202}
  {\bibfield  {journal} {\bibinfo  {journal} {Physical Review Research}\
  }\textbf {\bibinfo {volume} {4}},\ \bibinfo {pages} {023202} (\bibinfo {year}
  {2022})}\BibitemShut {NoStop}%
\bibitem [{\citenamefont {Zirnbauer}()}]{Zirnbauer1999}%
  \BibitemOpen
  \bibfield  {author} {\bibinfo {author} {\bibfnamefont {M.~R.}\ \bibnamefont
  {Zirnbauer}},\ }\href@noop {} {\enquote {\bibinfo {title} {Conformal field
  theory of the integer quantum {Hall} plateau transition},}\ }\bibinfo
  {howpublished} {arXiv:hep-th/9905054}\BibitemShut {NoStop}%
\bibitem [{\citenamefont {Kettemann}\ and\ \citenamefont
  {Tsvelik}(1999)}]{Kettemann1999}%
  \BibitemOpen
  \bibfield  {author} {\bibinfo {author} {\bibfnamefont {S.}~\bibnamefont
  {Kettemann}}\ and\ \bibinfo {author} {\bibfnamefont {A.~M.}\ \bibnamefont
  {Tsvelik}},\ }\bibfield  {title} {\enquote {\bibinfo {title} {Information
  about the integer quantum {Hall} transition extracted from the
  autocorrelation function of spectral determinants},}\ }\href {\doibase
  10.1103/PhysRevLett.82.3689} {\bibfield  {journal} {\bibinfo  {journal}
  {Phys. Rev. Lett.}\ }\textbf {\bibinfo {volume} {82}},\ \bibinfo {pages}
  {3689} (\bibinfo {year} {1999})}\BibitemShut {NoStop}%
\bibitem [{\citenamefont {Bhaseen}\ \emph {et~al.}(2000)\citenamefont
  {Bhaseen}, \citenamefont {Kogan}, \citenamefont {Soloviev}, \citenamefont
  {Taniguchi},\ and\ \citenamefont {Tsvelik}}]{Bhaseen2000}%
  \BibitemOpen
  \bibfield  {author} {\bibinfo {author} {\bibfnamefont {M.~J.}\ \bibnamefont
  {Bhaseen}}, \bibinfo {author} {\bibfnamefont {I.~I.}\ \bibnamefont {Kogan}},
  \bibinfo {author} {\bibfnamefont {O.~A.}\ \bibnamefont {Soloviev}}, \bibinfo
  {author} {\bibfnamefont {N.}~\bibnamefont {Taniguchi}}, \ and\ \bibinfo
  {author} {\bibfnamefont {A.~M.}\ \bibnamefont {Tsvelik}},\ }\bibfield
  {title} {\enquote {\bibinfo {title} {Towards a field theory of the plateau
  transitions in the integer quantum {Hall} effect},}\ }\href {\doibase
  https://doi.org/10.1016/S0550-3213(00)00276-5} {\bibfield  {journal}
  {\bibinfo  {journal} {Nucl. Phys. B}\ }\textbf {\bibinfo {volume} {580}},\
  \bibinfo {pages} {688} (\bibinfo {year} {2000})}\BibitemShut {NoStop}%
\bibitem [{\citenamefont {Tsvelik}()}]{Tsvelik2001}%
  \BibitemOpen
  \bibfield  {author} {\bibinfo {author} {\bibfnamefont {A.~M.}\ \bibnamefont
  {Tsvelik}},\ }\href@noop {} {\enquote {\bibinfo {title} {Wave functions
  statistics at quantum hall critical point},}\ }\bibinfo {howpublished}
  {arXiv:cond-mat/0112008}\BibitemShut {NoStop}%
\bibitem [{\citenamefont {Tsvelik}(2007)}]{Tsvelik2007}%
  \BibitemOpen
  \bibfield  {author} {\bibinfo {author} {\bibfnamefont {A.~M.}\ \bibnamefont
  {Tsvelik}},\ }\bibfield  {title} {\enquote {\bibinfo {title} {Evidence for
  the {PSL(2|2) Wess-Zumino-Novikov-Witten} model as a model for the plateau
  transition in the quantum {Hall} effect: {Evaluation} of numerical
  simulations},}\ }\href {\doibase 10.1103/PhysRevB.75.184201} {\bibfield
  {journal} {\bibinfo  {journal} {Phys. Rev. B}\ }\textbf {\bibinfo {volume}
  {75}},\ \bibinfo {pages} {184201} (\bibinfo {year} {2007})}\BibitemShut
  {NoStop}%
\bibitem [{\citenamefont {Zirnbauer}(2019)}]{Zirnbauer2019}%
  \BibitemOpen
  \bibfield  {author} {\bibinfo {author} {\bibfnamefont {M.~R.}\ \bibnamefont
  {Zirnbauer}},\ }\bibfield  {title} {\enquote {\bibinfo {title} {The integer
  quantum {Hall} plateau transition is a current algebra after all},}\ }\href
  {\doibase https://doi.org/10.1016/j.nuclphysb.2019.02.017} {\bibfield
  {journal} {\bibinfo  {journal} {Nucl. Phys. B}\ }\textbf {\bibinfo {volume}
  {941}},\ \bibinfo {pages} {458} (\bibinfo {year} {2019})}\BibitemShut
  {NoStop}%
\bibitem [{\citenamefont {Padayasi}\ and\ \citenamefont
  {Gruzberg}()}]{Padayasi2023}%
  \BibitemOpen
  \bibfield  {author} {\bibinfo {author} {\bibfnamefont {Jaychandran}\
  \bibnamefont {Padayasi}}\ and\ \bibinfo {author} {\bibfnamefont {Ilya~A.}\
  \bibnamefont {Gruzberg}},\ }\href@noop {} {\enquote {\bibinfo {title}
  {Conformal invariance and multifractality at {Anderson} transitions in
  arbitrary dimensions},}\ }\bibinfo {howpublished} {arXiv:
  2306.07340}\BibitemShut {NoStop}%
\bibitem [{\citenamefont {Bondesan}\ \emph {et~al.}(2017)\citenamefont
  {Bondesan}, \citenamefont {Wieczorek},\ and\ \citenamefont
  {Zirnbauer}}]{Bondesan2017}%
  \BibitemOpen
  \bibfield  {author} {\bibinfo {author} {\bibfnamefont {R.}~\bibnamefont
  {Bondesan}}, \bibinfo {author} {\bibfnamefont {D.}~\bibnamefont {Wieczorek}},
  \ and\ \bibinfo {author} {\bibfnamefont {M.R.}\ \bibnamefont {Zirnbauer}},\
  }\bibfield  {title} {\enquote {\bibinfo {title} {Gaussian free fields at the
  integer quantum {Hall} plateau transition},}\ }\href {\doibase
  https://doi.org/10.1016/j.nuclphysb.2017.02.011} {\bibfield  {journal}
  {\bibinfo  {journal} {Nucl. Phys. B}\ }\textbf {\bibinfo {volume} {918}},\
  \bibinfo {pages} {52} (\bibinfo {year} {2017})}\BibitemShut {NoStop}%
\bibitem [{\citenamefont {Kagalovsky}\ \emph {et~al.}(1999)\citenamefont
  {Kagalovsky}, \citenamefont {Horovitz}, \citenamefont {Avishai},\ and\
  \citenamefont {Chalker}}]{Kagolovsky1999}%
  \BibitemOpen
  \bibfield  {author} {\bibinfo {author} {\bibfnamefont {V.}~\bibnamefont
  {Kagalovsky}}, \bibinfo {author} {\bibfnamefont {B.}~\bibnamefont
  {Horovitz}}, \bibinfo {author} {\bibfnamefont {Y.}~\bibnamefont {Avishai}}, \
  and\ \bibinfo {author} {\bibfnamefont {J.~T.}\ \bibnamefont {Chalker}},\
  }\bibfield  {title} {\enquote {\bibinfo {title} {Quantum {Hall} plateau
  transitions in disordered superconductors},}\ }\href {\doibase
  10.1103/PhysRevLett.82.3516} {\bibfield  {journal} {\bibinfo  {journal}
  {Phys. Rev. Lett.}\ }\textbf {\bibinfo {volume} {82}},\ \bibinfo {pages}
  {3516} (\bibinfo {year} {1999})}\BibitemShut {NoStop}%
\bibitem [{\citenamefont {Senthil}\ \emph {et~al.}(1999)\citenamefont
  {Senthil}, \citenamefont {Marston},\ and\ \citenamefont
  {Fisher}}]{Senthil1999}%
  \BibitemOpen
  \bibfield  {author} {\bibinfo {author} {\bibfnamefont {T.}~\bibnamefont
  {Senthil}}, \bibinfo {author} {\bibfnamefont {J.~B.}\ \bibnamefont
  {Marston}}, \ and\ \bibinfo {author} {\bibfnamefont {Matthew P.~A.}\
  \bibnamefont {Fisher}},\ }\bibfield  {title} {\enquote {\bibinfo {title}
  {Spin quantum hall effect in unconventional superconductors},}\ }\href
  {\doibase 10.1103/PhysRevB.60.4245} {\bibfield  {journal} {\bibinfo
  {journal} {Phys. Rev. B}\ }\textbf {\bibinfo {volume} {60}},\ \bibinfo
  {pages} {4245} (\bibinfo {year} {1999})}\BibitemShut {NoStop}%
\bibitem [{\citenamefont {Gruzberg}\ \emph {et~al.}(1999)\citenamefont
  {Gruzberg}, \citenamefont {Ludwig},\ and\ \citenamefont
  {Read}}]{Gruzberg1999}%
  \BibitemOpen
  \bibfield  {author} {\bibinfo {author} {\bibfnamefont {Ilya~A.}\ \bibnamefont
  {Gruzberg}}, \bibinfo {author} {\bibfnamefont {Andreas W.~W.}\ \bibnamefont
  {Ludwig}}, \ and\ \bibinfo {author} {\bibfnamefont {N.}~\bibnamefont
  {Read}},\ }\bibfield  {title} {\enquote {\bibinfo {title} {Exact exponents
  for the spin quantum {Hall} transition},}\ }\href {\doibase
  10.1103/PhysRevLett.82.4524} {\bibfield  {journal} {\bibinfo  {journal}
  {Phys. Rev. Lett.}\ }\textbf {\bibinfo {volume} {82}},\ \bibinfo {pages}
  {4524} (\bibinfo {year} {1999})}\BibitemShut {NoStop}%
\bibitem [{\citenamefont {Beamond}\ \emph {et~al.}(2002)\citenamefont
  {Beamond}, \citenamefont {Cardy},\ and\ \citenamefont
  {Chalker}}]{Beamond2002}%
  \BibitemOpen
  \bibfield  {author} {\bibinfo {author} {\bibfnamefont {E.~J.}\ \bibnamefont
  {Beamond}}, \bibinfo {author} {\bibfnamefont {John}\ \bibnamefont {Cardy}}, \
  and\ \bibinfo {author} {\bibfnamefont {J.~T.}\ \bibnamefont {Chalker}},\
  }\bibfield  {title} {\enquote {\bibinfo {title} {Quantum and classical
  localization, the spin quantum {Hall} effect, and generalizations},}\ }\href
  {\doibase 10.1103/PhysRevB.65.214301} {\bibfield  {journal} {\bibinfo
  {journal} {Phys. Rev. B}\ }\textbf {\bibinfo {volume} {65}},\ \bibinfo
  {pages} {214301} (\bibinfo {year} {2002})}\BibitemShut {NoStop}%
\bibitem [{\citenamefont {Mirlin}\ \emph {et~al.}(2003)\citenamefont {Mirlin},
  \citenamefont {Evers},\ and\ \citenamefont {Mildenberger}}]{Mirlin2003}%
  \BibitemOpen
  \bibfield  {author} {\bibinfo {author} {\bibfnamefont {A.~D.}\ \bibnamefont
  {Mirlin}}, \bibinfo {author} {\bibfnamefont {F.}~\bibnamefont {Evers}}, \
  and\ \bibinfo {author} {\bibfnamefont {A.}~\bibnamefont {Mildenberger}},\
  }\bibfield  {title} {\enquote {\bibinfo {title} {Wavefunction statistics and
  multifractality at the spin quantum {Hall} transition},}\ }\href {\doibase
  10.1088/0305-4470/36/12/323} {\bibfield  {journal} {\bibinfo  {journal} {J.
  Phys. A: Math. and Gen.}\ }\textbf {\bibinfo {volume} {36}},\ \bibinfo
  {pages} {3255} (\bibinfo {year} {2003})}\BibitemShut {NoStop}%
\bibitem [{\citenamefont {Evers}\ \emph {et~al.}(2003)\citenamefont {Evers},
  \citenamefont {Mildenberger},\ and\ \citenamefont {Mirlin}}]{Evers2003}%
  \BibitemOpen
  \bibfield  {author} {\bibinfo {author} {\bibfnamefont {F.}~\bibnamefont
  {Evers}}, \bibinfo {author} {\bibfnamefont {A.}~\bibnamefont {Mildenberger}},
  \ and\ \bibinfo {author} {\bibfnamefont {A.~D.}\ \bibnamefont {Mirlin}},\
  }\bibfield  {title} {\enquote {\bibinfo {title} {Multifractality at the spin
  quantum {Hall} transition},}\ }\href {\doibase 10.1103/PhysRevB.67.041303}
  {\bibfield  {journal} {\bibinfo  {journal} {Phys. Rev. B}\ }\textbf {\bibinfo
  {volume} {67}},\ \bibinfo {pages} {041303} (\bibinfo {year}
  {2003})}\BibitemShut {NoStop}%
\bibitem [{\citenamefont {Puschmann}\ \emph {et~al.}(2021)\citenamefont
  {Puschmann}, \citenamefont {Hernang\'omez-P\'erez}, \citenamefont {Lang},
  \citenamefont {Bera},\ and\ \citenamefont {Evers}}]{Puschmann2021}%
  \BibitemOpen
  \bibfield  {author} {\bibinfo {author} {\bibfnamefont {Martin}\ \bibnamefont
  {Puschmann}}, \bibinfo {author} {\bibfnamefont {Daniel}\ \bibnamefont
  {Hernang\'omez-P\'erez}}, \bibinfo {author} {\bibfnamefont {Bruno}\
  \bibnamefont {Lang}}, \bibinfo {author} {\bibfnamefont {Soumya}\ \bibnamefont
  {Bera}}, \ and\ \bibinfo {author} {\bibfnamefont {Ferdinand}\ \bibnamefont
  {Evers}},\ }\bibfield  {title} {\enquote {\bibinfo {title} {Quartic
  multifractality and finite-size corrections at the spin quantum {Hall}
  transition},}\ }\href {\doibase 10.1103/PhysRevB.103.235167} {\bibfield
  {journal} {\bibinfo  {journal} {Phys. Rev. B}\ }\textbf {\bibinfo {volume}
  {103}},\ \bibinfo {pages} {235167} (\bibinfo {year} {2021})}\BibitemShut
  {NoStop}%
\bibitem [{\citenamefont {Finkel'stein}(1990)}]{Fin}%
  \BibitemOpen
  \bibfield  {author} {\bibinfo {author} {\bibfnamefont {A.~M.}\ \bibnamefont
  {Finkel'stein}},\ }\bibfield  {title} {\enquote {\bibinfo {title} {Electron
  liquid in disordered conductors},}\ }in\ \href@noop {} {\emph {\bibinfo
  {booktitle} {Soviet Scientific Reviews}}},\ Vol.~\bibinfo {volume} {14},\
  \bibinfo {editor} {edited by\ \bibinfo {editor} {\bibfnamefont {I.~M.}\
  \bibnamefont {Khalatnikov}}}\ (\bibinfo  {publisher} {Harwood Academic
  Publishers, London},\ \bibinfo {year} {1990})\BibitemShut {NoStop}%
\bibitem [{\citenamefont {Belitz}\ and\ \citenamefont
  {Kirkpatrick}(1994)}]{Belitz1994}%
  \BibitemOpen
  \bibfield  {author} {\bibinfo {author} {\bibfnamefont {D.}~\bibnamefont
  {Belitz}}\ and\ \bibinfo {author} {\bibfnamefont {T.~R.}\ \bibnamefont
  {Kirkpatrick}},\ }\bibfield  {title} {\enquote {\bibinfo {title} {The
  {Anderson-Mott} transition},}\ }\href {\doibase 10.1103/RevModPhys.66.261}
  {\bibfield  {journal} {\bibinfo  {journal} {Rev. Mod. Phys.}\ }\textbf
  {\bibinfo {volume} {66}},\ \bibinfo {pages} {261} (\bibinfo {year}
  {1994})}\BibitemShut {NoStop}%
\bibitem [{\citenamefont {Burmistrov}\ \emph {et~al.}(2013)\citenamefont
  {Burmistrov}, \citenamefont {Gornyi},\ and\ \citenamefont
  {Mirlin}}]{Burmistrov2013}%
  \BibitemOpen
  \bibfield  {author} {\bibinfo {author} {\bibfnamefont {I.~S.}\ \bibnamefont
  {Burmistrov}}, \bibinfo {author} {\bibfnamefont {I.~V.}\ \bibnamefont
  {Gornyi}}, \ and\ \bibinfo {author} {\bibfnamefont {A.~D.}\ \bibnamefont
  {Mirlin}},\ }\bibfield  {title} {\enquote {\bibinfo {title} {Multifractality
  at {Anderson} transitions with {Coulomb} interaction},}\ }\href {\doibase
  10.1103/PhysRevLett.111.066601} {\bibfield  {journal} {\bibinfo  {journal}
  {Phys. Rev. Lett.}\ }\textbf {\bibinfo {volume} {111}},\ \bibinfo {pages}
  {066601} (\bibinfo {year} {2013})}\BibitemShut {NoStop}%
\bibitem [{\citenamefont {Burmistrov}\ \emph
  {et~al.}(2015{\natexlab{b}})\citenamefont {Burmistrov}, \citenamefont
  {Gornyi},\ and\ \citenamefont {Mirlin}}]{Burmistrov2015m}%
  \BibitemOpen
  \bibfield  {author} {\bibinfo {author} {\bibfnamefont {I.~S.}\ \bibnamefont
  {Burmistrov}}, \bibinfo {author} {\bibfnamefont {I.~V.}\ \bibnamefont
  {Gornyi}}, \ and\ \bibinfo {author} {\bibfnamefont {A.~D.}\ \bibnamefont
  {Mirlin}},\ }\bibfield  {title} {\enquote {\bibinfo {title} {Multifractality
  and electron-electron interaction at {Anderson} transitions},}\ }\href
  {\doibase 10.1103/PhysRevB.91.085427} {\bibfield  {journal} {\bibinfo
  {journal} {Phys. Rev. B}\ }\textbf {\bibinfo {volume} {91}},\ \bibinfo
  {pages} {085427} (\bibinfo {year} {2015}{\natexlab{b}})}\BibitemShut
  {NoStop}%
\bibitem [{\citenamefont {Repin}\ and\ \citenamefont
  {Burmistrov}(2016)}]{Repin2016}%
  \BibitemOpen
  \bibfield  {author} {\bibinfo {author} {\bibfnamefont {E.~V.}\ \bibnamefont
  {Repin}}\ and\ \bibinfo {author} {\bibfnamefont {I.~S.}\ \bibnamefont
  {Burmistrov}},\ }\bibfield  {title} {\enquote {\bibinfo {title} {Mesoscopic
  fluctuations of the single-particle {Green's} function at {Anderson}
  transitions with {Coulomb} interaction},}\ }\href {\doibase
  10.1103/PhysRevB.94.245442} {\bibfield  {journal} {\bibinfo  {journal} {Phys.
  Rev. B}\ }\textbf {\bibinfo {volume} {94}},\ \bibinfo {pages} {245442}
  (\bibinfo {year} {2016})}\BibitemShut {NoStop}%
\bibitem [{\citenamefont {Babkin}\ and\ \citenamefont
  {Burmistrov}(2022)}]{Babkin2022}%
  \BibitemOpen
  \bibfield  {author} {\bibinfo {author} {\bibfnamefont {S.~S.}\ \bibnamefont
  {Babkin}}\ and\ \bibinfo {author} {\bibfnamefont {I.~S.}\ \bibnamefont
  {Burmistrov}},\ }\bibfield  {title} {\enquote {\bibinfo {title} {Generalized
  multifractality in the spin quantum hall symmetry class with interaction},}\
  }\href {\doibase 10.1103/PhysRevB.106.125424} {\bibfield  {journal} {\bibinfo
   {journal} {Phys. Rev. B}\ }\textbf {\bibinfo {volume} {106}},\ \bibinfo
  {pages} {125424} (\bibinfo {year} {2022})}\BibitemShut {NoStop}%
\bibitem [{\citenamefont {Amit}(1993)}]{Amit-book}%
  \BibitemOpen
  \bibfield  {author} {\bibinfo {author} {\bibfnamefont {D.~J.}\ \bibnamefont
  {Amit}},\ }\bibfield  {title} {\enquote {\bibinfo {title} {{Field Theory, the
  Renormalization Group, and Critical Phenomena}},}\ \ }(\bibinfo  {publisher}
  {World Scientific, Singapore},\ \bibinfo {year} {1993})\BibitemShut {NoStop}%
\bibitem [{\citenamefont {Jeng}\ \emph
  {et~al.}(2001{\natexlab{a}})\citenamefont {Jeng}, \citenamefont {Ludwig},
  \citenamefont {Senthil},\ and\ \citenamefont {Chamon}}]{Jeng2001a}%
  \BibitemOpen
  \bibfield  {author} {\bibinfo {author} {\bibfnamefont {M.}~\bibnamefont
  {Jeng}}, \bibinfo {author} {\bibfnamefont {A.~W.~W.}\ \bibnamefont {Ludwig}},
  \bibinfo {author} {\bibfnamefont {T.}~\bibnamefont {Senthil}}, \ and\
  \bibinfo {author} {\bibfnamefont {C.}~\bibnamefont {Chamon}},\ }\bibfield
  {title} {\enquote {\bibinfo {title} {Interaction effects on quasiparticle
  localization in dirty superconductors},}\ }\href@noop {} {\bibfield
  {journal} {\bibinfo  {journal} {Bull. Am. Phys. Soc.}\ }\textbf {\bibinfo
  {volume} {46}},\ \bibinfo {pages} {231} (\bibinfo {year}
  {2001}{\natexlab{a}})}\BibitemShut {NoStop}%
\bibitem [{\citenamefont {Jeng}\ \emph
  {et~al.}(2001{\natexlab{b}})\citenamefont {Jeng}, \citenamefont {Ludwig},
  \citenamefont {Senthil},\ and\ \citenamefont {Chamon}}]{Jeng2001}%
  \BibitemOpen
  \bibfield  {author} {\bibinfo {author} {\bibfnamefont {M.}~\bibnamefont
  {Jeng}}, \bibinfo {author} {\bibfnamefont {A.~W.~W.}\ \bibnamefont {Ludwig}},
  \bibinfo {author} {\bibfnamefont {T.}~\bibnamefont {Senthil}}, \ and\
  \bibinfo {author} {\bibfnamefont {C.}~\bibnamefont {Chamon}},\ }\href@noop {}
  {\enquote {\bibinfo {title} {Interaction effects on quasiparticle
  localization in dirty superconductors},}\ }\bibinfo {howpublished}
  {arXiv:cond-mat/0112044} (\bibinfo {year} {2001}{\natexlab{b}})\BibitemShut
  {NoStop}%
\bibitem [{\citenamefont {Dell'Anna}(2006)}]{DellAnna2006}%
  \BibitemOpen
  \bibfield  {author} {\bibinfo {author} {\bibfnamefont {Luca}\ \bibnamefont
  {Dell'Anna}},\ }\bibfield  {title} {\enquote {\bibinfo {title} {Disordered
  d-wave superconductors with interactions},}\ }\href {\doibase
  https://doi.org/10.1016/j.nuclphysb.2006.09.024} {\bibfield  {journal}
  {\bibinfo  {journal} {Nucl. Phys. B}\ }\textbf {\bibinfo {volume} {758}},\
  \bibinfo {pages} {255} (\bibinfo {year} {2006})}\BibitemShut {NoStop}%
\bibitem [{\citenamefont {Liao}\ \emph {et~al.}(2017)\citenamefont {Liao},
  \citenamefont {Levchenko},\ and\ \citenamefont {Foster}}]{Liao2017}%
  \BibitemOpen
  \bibfield  {author} {\bibinfo {author} {\bibfnamefont {Yunxiang}\
  \bibnamefont {Liao}}, \bibinfo {author} {\bibfnamefont {Alex}\ \bibnamefont
  {Levchenko}}, \ and\ \bibinfo {author} {\bibfnamefont {Matthew~S.}\
  \bibnamefont {Foster}},\ }\bibfield  {title} {\enquote {\bibinfo {title}
  {Response theory of the ergodic many-body delocalized phase: {Keldysh
  Finkel'stein} sigma models and the 10-fold way},}\ }\href {\doibase
  https://doi.org/10.1016/j.aop.2017.08.020} {\bibfield  {journal} {\bibinfo
  {journal} {Ann. Phys. (N.Y.)}\ }\textbf {\bibinfo {volume} {386}},\ \bibinfo
  {pages} {97} (\bibinfo {year} {2017})}\BibitemShut {NoStop}%
\bibitem [{\citenamefont {Pruisken}\ and\ \citenamefont
  {Burmistrov}(2005)}]{Pruisken2005}%
  \BibitemOpen
  \bibfield  {author} {\bibinfo {author} {\bibfnamefont {A.M.M.}\ \bibnamefont
  {Pruisken}}\ and\ \bibinfo {author} {\bibfnamefont {I.S.}\ \bibnamefont
  {Burmistrov}},\ }\bibfield  {title} {\enquote {\bibinfo {title} {The
  instanton vacuum of generalized {CP$^{N-1}$} models},}\ }\href {\doibase
  https://doi.org/10.1016/j.aop.2004.08.009} {\bibfield  {journal} {\bibinfo
  {journal} {Ann. Phys. (N.Y.)}\ }\textbf {\bibinfo {volume} {316}},\ \bibinfo
  {pages} {285} (\bibinfo {year} {2005})}\BibitemShut {NoStop}%
\bibitem [{\citenamefont {Harashima}\ and\ \citenamefont
  {Slevin}(2012)}]{Slevin2012}%
  \BibitemOpen
  \bibfield  {author} {\bibinfo {author} {\bibfnamefont {Y.}~\bibnamefont
  {Harashima}}\ and\ \bibinfo {author} {\bibfnamefont {K.}~\bibnamefont
  {Slevin}},\ }\bibfield  {title} {\enquote {\bibinfo {title} {Effect of
  electron-electron interaction near the metal-insulator transition in doped
  semiconductors studied within the local density approximation},}\ }\href
  {http://www.worldscientific.com/doi/abs/10.1142/S2010194512005958} {\bibfield
   {journal} {\bibinfo  {journal} {Int. J. Mod. Phys. Conf. Ser.}\ }\textbf
  {\bibinfo {volume} {11}},\ \bibinfo {pages} {90} (\bibinfo {year}
  {2012})}\BibitemShut {NoStop}%
\bibitem [{\citenamefont {Slevin}\ and\ \citenamefont
  {Ohtsuki}(2014)}]{Slevin2014}%
  \BibitemOpen
  \bibfield  {author} {\bibinfo {author} {\bibfnamefont {K.}~\bibnamefont
  {Slevin}}\ and\ \bibinfo {author} {\bibfnamefont {T.}~\bibnamefont
  {Ohtsuki}},\ }\bibfield  {title} {\enquote {\bibinfo {title} {Critical
  exponent for the {Anderson} transition in the three-dimensional orthogonal
  universality class},}\ }\href
  {http://stacks.iop.org/1367-2630/16/i=1/a=015012} {\bibfield  {journal}
  {\bibinfo  {journal} {New J. Phys.}\ }\textbf {\bibinfo {volume} {16}},\
  \bibinfo {pages} {015012} (\bibinfo {year} {2014})}\BibitemShut {NoStop}%
\bibitem [{\citenamefont {Amini}\ \emph {et~al.}(2014)\citenamefont {Amini},
  \citenamefont {Kravtsov},\ and\ \citenamefont {M{\"u}ller}}]{Amini2014}%
  \BibitemOpen
  \bibfield  {author} {\bibinfo {author} {\bibfnamefont {M.}~\bibnamefont
  {Amini}}, \bibinfo {author} {\bibfnamefont {V.~E.}\ \bibnamefont {Kravtsov}},
  \ and\ \bibinfo {author} {\bibfnamefont {M.}~\bibnamefont {M{\"u}ller}},\
  }\bibfield  {title} {\enquote {\bibinfo {title} {Multifractality and
  quantum-to-classical crossover in the {Coulomb} anomaly at the
  {Mott--Anderson} metal--insulator transition},}\ }\href
  {http://iopscience.iop.org/article/10.1088/1367-2630/16/1/015022/meta}
  {\bibfield  {journal} {\bibinfo  {journal} {New J. Phys.}\ }\textbf {\bibinfo
  {volume} {16}},\ \bibinfo {pages} {015022} (\bibinfo {year}
  {2014})}\BibitemShut {NoStop}%
\bibitem [{\citenamefont {Lee}\ and\ \citenamefont {Kim}(2018)}]{Lee2018}%
  \BibitemOpen
  \bibfield  {author} {\bibinfo {author} {\bibfnamefont {H.-J.}\ \bibnamefont
  {Lee}}\ and\ \bibinfo {author} {\bibfnamefont {K.-S.}\ \bibnamefont {Kim}},\
  }\bibfield  {title} {\enquote {\bibinfo {title} {{Hartree-Fock study of the
  Anderson metal-insulator transition in the presence of Coulomb interaction:
  Two types of mobility edges and their multifractal scaling exponents}},}\
  }\href {\doibase 10.1103/PhysRevB.97.155105} {\bibfield  {journal} {\bibinfo
  {journal} {Phys. Rev. B}\ }\textbf {\bibinfo {volume} {97}},\ \bibinfo
  {pages} {155105} (\bibinfo {year} {2018})}\BibitemShut {NoStop}%
\end{thebibliography}%
	
\end{document}